\documentclass[aps,prl,reprint,amsmath,amssymb,superscriptaddress,twocolumn,british]{revtex4-2}
\usepackage{times}
\usepackage{physics}
\newcommand{\prlsection}[1]{ \noindent\textbf{\textit{#1---}}}
\usepackage[protrusion=true,expansion=true]{microtype}
\newcommand{\be}{\begin{equation}}
\newcommand{\ee}{\end{equation}}
\newcommand{\de}{\partial}
\newcommand{\n}{\mathfrak{n}}
\newcommand{\jj}{\mathfrak{J}}
\usepackage{ulem}
\usepackage[T1]{fontenc}
\usepackage{overpic}
\usepackage{xcolor}
\usepackage[utf8]{inputenc}
\usepackage{babel}
\usepackage{verbatim}
\usepackage{amsmath}
\usepackage{amsthm}
\usepackage{amssymb}
\usepackage{graphicx}
\usepackage{tikz}

\DeclareRobustCommand{\opencircle}{%
  \tikz[baseline=-0.55ex]
  \draw[black,fill=none,line width=0.5pt]
  (0,0) circle[radius=0.78ex];%
}

\DeclareRobustCommand{\openstar}{%
  \tikz[baseline=-0.55ex,x=0.85ex,y=0.85ex]
  \draw[black,fill=none,line width=0.5pt]
  (90:1)
  -- (54:0.382)
  -- (18:1)
  -- (-18:0.382)
  -- (-54:1)
  -- (-90:0.382)
  -- (-126:1)
  -- (-162:0.382)
  -- (162:1)
  -- (126:0.382)
  -- cycle;%
}
\usepackage[pdfusetitle,bookmarks=true,bookmarksnumbered=false,bookmarksopen=false,breaklinks=false,pdfborder={0 0 1},backref=false,
    colorlinks=true,
    citecolor=blue,
    linkcolor=blue,
    urlcolor=blue
]{hyperref}

\newcommand{\alexios}[1]{\textcolor{blue}{#1}}

\definecolor{myorange}{HTML}{FDB658}

\newcommand{\titleinfo}{
Nonlinear Fluctuating Hydrodynamics from Interacting Noisy Quantum Matter}
\begin{document}
\title{\titleinfo}

\author{Alexios Christopoulos}
\thanks{Equal contributions. AC performed all numerical TEBD2 simulations of the microscopic model. JC derived the effective ISEP dynamics from the microscopic model in the diffusive strong-coupling regime and contributed to the computations of the NESS of the ISEP. SS established the MFT framework and carried out the calculations of the density profile, correlations, and full counting statistics (FCS) of the ISEP model both from the microscopic definition of the model and the MFT.}
\affiliation{Jožef Stefan Institute, 1000 Ljubljana, Slovenia}

\author{Jo\~ao Costa}
\thanks{Equal contributions. AC performed all numerical TEBD2 simulations of the microscopic model. JC derived the effective ISEP dynamics from the microscopic model in the diffusive strong-coupling regime and contributed to the computations of the NESS of the ISEP. SS established the MFT framework and carried out the calculations of the density profile, correlations, and full counting statistics (FCS) of the ISEP model both from the microscopic definition of the model and the MFT.}
\affiliation{CeFEMA-LaPMET, Departamento de Física, Instituto Superior Técnico, Universidade de Lisboa, Av. Rovisco Pais, 1049-001 Lisboa, Portugal}
\affiliation{Laboratoire de Physique Th\'eorique et Mod\'elisation, CNRS UMR 8089, CY Cergy Paris Universit\'e, 95302 Cergy-Pontoise Cedex, France}

\author{Stefano Scopa}
\thanks{Equal contributions. AC performed all numerical TEBD2 simulations of the microscopic model. JC derived the effective ISEP dynamics from the microscopic model in the diffusive strong-coupling regime and contributed to the computations of the NESS of the ISEP. SS established the MFT framework and carried out the calculations of the density profile, correlations, and full counting statistics (FCS) of the ISEP model both from the microscopic definition of the model and the MFT.}
\affiliation{Laboratoire de Physique de l’\'Ecole Normale Sup\'erieure, CNRS,
ENS \& Universit\'e PSL, Sorbonne Universit\'e, Universit\'e Paris Cit\'e, 75005 Paris, France.}

\author{Jacopo de Nardis}
\affiliation{Laboratoire de Physique Th\'eorique et Mod\'elisation, CNRS UMR 8089, CY Cergy Paris Universit\'e, 95302 Cergy-Pontoise Cedex, France}
\affiliation{JEIP, UAR 3573 CNRS, Collège de France, PSL Research University,
11 Place Marcelin Berthelot, 75321 Paris Cedex 05, France}

\author{Zala Lenar\v{c}i\v{c}}
\affiliation{Jožef Stefan Institute, 1000 Ljubljana, Slovenia}

\author{Denis Bernard}
\affiliation{Laboratoire de Physique de l’\'Ecole Normale Sup\'erieure, CNRS,
ENS \& Universit\'e PSL, Sorbonne Universit\'e, Universit\'e Paris Cit\'e, 75005 Paris, France.}

\author{Tony Jin}
\email{tony.jin@univ-cotedazur.fr}
\affiliation{Universit\'e C\^ote d'Azur, CNRS, Centrale Med, Institut de Physique de Nice, 06200 Nice, France}

\begin{abstract}
A universal characterization of non-equilibrium steady states in interacting quantum many-body systems remains one of the central challenges of statistical physics. Here, we address this problem for a paradigmatic model of diffusive interacting quantum matter---the boundary-driven XXZ spin chain with bulk dephasing---and derive, directly from its microscopic Lindblad dynamics, an emergent classical Macroscopic Fluctuation Theory (MFT) governing its large-scale fluctuations. Crucially, the resulting hydrodynamics carries a density-dependent diffusivity and mobility as the fingerprint of interactions. This effective description enables the exact computation of the stationary density profile, long-range correlations, and the full counting statistics of the current, in excellent agreement with tensor-network simulations. Our work demonstrates that noisy quantum many-body systems can realize the universality class of genuinely interacting diffusive matter, beyond the constant-diffusivity class of the symmetric simple exclusion process, and establishes MFT as a powerful universal framework for interacting diffusive quantum systems.
\end{abstract}
\maketitle

The success of equilibrium statistical mechanics rests on the existence of universal ensembles that reduce the description of a macroscopic system to a handful of thermodynamic parameters. Whether an equally universal characterization exists for matter driven \emph{out of equilibrium} remains one of the central open problems of statistical physics~\cite{derrida_2007}. For classical
diffusive systems, such a framework exists: The Macroscopic Fluctuation Theory (MFT)~\cite{Bertini2005,bertini2001fluctuations,bertini2002macroscopic,mft2015rmp} provides a fluctuating hydrodynamic description of their large deviations that depends only on two transport coefficients, the diffusivity and the mobility. Much like Gibbs ensembles at equilibrium, the MFT captures the macroscopic physics independently of microscopic details.

Whether a comparably universal framework exists for interacting quantum many-body systems remains largely unknown. Nonequilibrium fluctuations inherit the exponential complexity of the underlying many-body problem, with no organizing principle analogous to the Gibbs ensemble. Thus, the central challenge is not merely to solve individual quantum models, but to uncover the universal hydrodynamic structures governing interacting quantum matter and understand how they emerge from microscopic quantum dynamics.

The first studies originate from the mesoscopic physics community with the computation of the  Full-Counting Statistics (FCS) of the current of disordered conductors~\cite{Beenaker1997_RMT_quantum_transport,LeeLevitovYuPRBUniversalstatistics,Hruza_Jin_QSSEP_Anderson}. Important progress has been made in recent years by describing the coupling to the environment through Markovian (GKLS) dynamics~\cite{Lindblad1976,Gorini1976}. Exact solutions have been determined for quadratic Liouvillians~\cite{Prosen_thirdquantization}, such as free fermions with local dephasing~\cite{Znidaric__XXdeph,znidaric2010njp,znidaric_2014_LD,ProsenEssler_Mapping,BauerBernardJin_Stoqdissipative,Garrahan_dephasing_boundary_driven,Jin_Quantumresistors,essler2026_dephasing,Sasamoto_2026_exactcurrentfluctuationstightbinding} and random hopping amplitudes~\cite{eisler2011,BernardJin_QSSEP,HruzaBernardPRX}. 
These works reveal an emerging universality: at the leading order in the thermodynamic limit, fluctuations of noisy quantum diffusive systems are described by the classical framework of MFT~\cite{LeeLevitovYuPRBUniversalstatistics,Abanin_Michailidis_XXZ_deph,Costa_emergence_universality_2026_PRL}, with genuinely quantum effects appearing only at subleading order in the large-deviation expansion~\cite{albert2026universalclassicalquantumfluctuations} or through intrinsically quantum observables such as entanglement~\cite{BernardPiroli_QSSEPentanglement,HruzaBenardEntanglementQSSEP,Bernard2026-domain}.

Despite this progress, 
analytically tractable descriptions of noisy quantum systems in terms of diffusive MFT have so far been largely restricted to density-independent diffusion, placing them in the same hydrodynamic universality class as the Symmetric Simple Exclusion Process (SSEP). 
A microscopic realization of nonlinear MFT, with density-dependent transport coefficients, has remained elusive, despite recent applications of MFT in interacting random circuits~\cite{FCSMFTDeNardis,SinghMcCulloch2025,McCulloch2026}. This gap is fundamental since a density-dependent diffusivity is the hydrodynamic fingerprint of interactions, underlying the rich phenomenology of interacting diffusive systems: nonlinear density profiles, long-range correlations, and non-Gaussian current fluctuations~\cite{HSpohn_1983,derrida_2007,mft2015rmp,Bodineau_2005,Baek_2017,SahaSadhu2026Density}. 

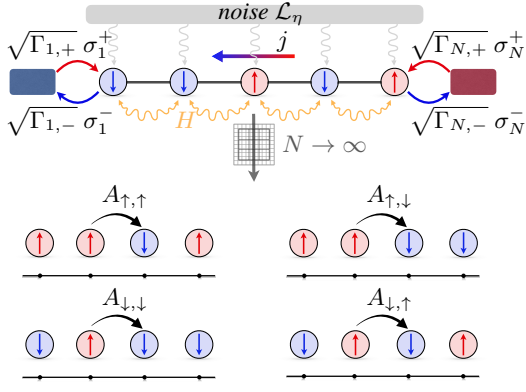
\begin{figure}
\centering
\begin{overpic}[width=0.75\columnwidth,trim={0.1cm 13cm 0.3cm 0.1cm},clip]{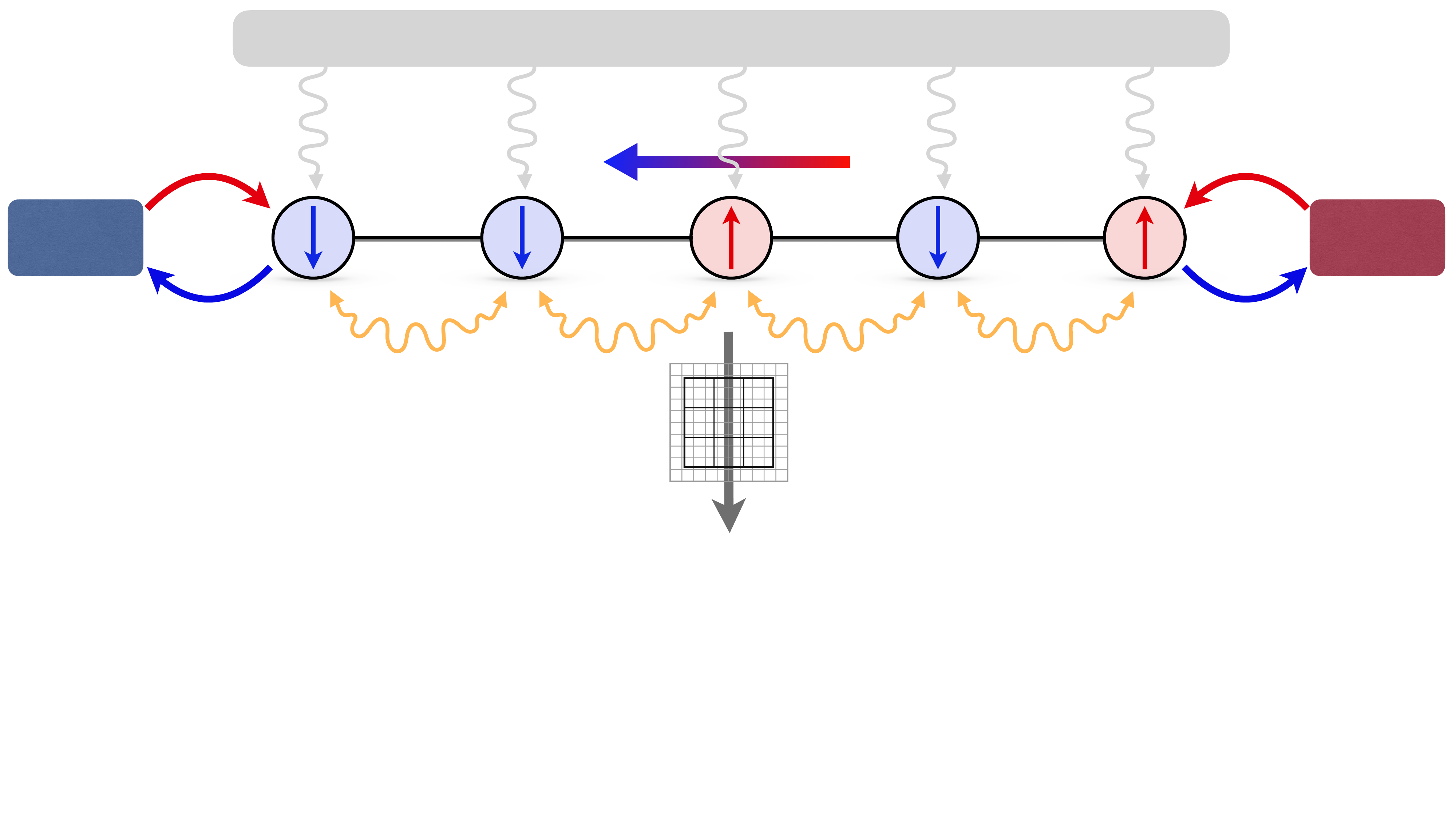}
\put(34,11.4){\textcolor{myorange}{$H$}}
\put(56,06.5){\textcolor{gray!130}{$N \to \infty$}}
\put(43,33.6){\textit{noise} $\mathcal{L}_{\eta}$}
\put(55,28){$j$}
\put(-0.5,27){$\sqrt{\Gamma_{1,+}}\ \sigma_1^+$}
\put(-0.5,11.5){$\sqrt{\Gamma_{1,-}}\ \sigma_1^-$}
\put(82.0,27){$\sqrt{\Gamma_{N,+}}\ \sigma_N^+$}
\put(82.0,11.5){$\sqrt{\Gamma_{N,-}}\ \sigma_N^-$}
\end{overpic}
   \par\vspace{3mm}
\begin{overpic}[width=0.75\columnwidth,trim={0.5cm 9.5cm 0.5cm 3.2cm},clip]{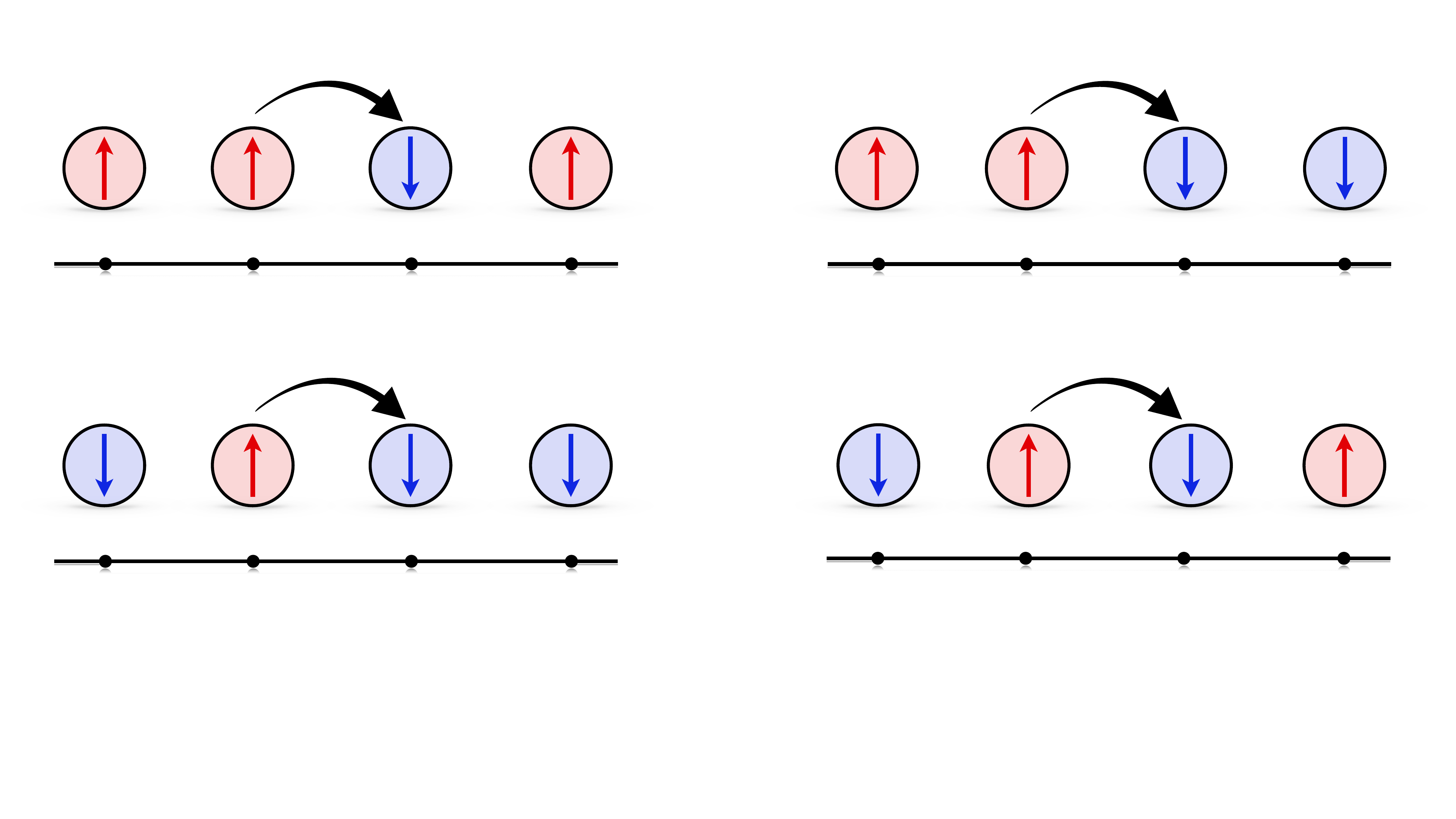}
\put(19.18,39.5){$A_{\uparrow,\uparrow}$}
\put(73.2,39.5){$A_{\uparrow,\downarrow}$}
\put(73.2,18.5){$A_{\downarrow,\uparrow}$}
\put(19.18,18.5){$A_{\downarrow,\downarrow}$}
\end{overpic}
\caption{\textbf{Microscopic model and emergent stochastic dynamics.} \textbf{Top:} XXZ spin chain with bulk dephasing and boundary reservoirs, driving the system to a NESS. \textbf{Bottom:} In the strong-coupling regime, the model reduces to the effective dynamics of the interacting symmetric exclusion process (ISEP), which describes non-overlapping particles whose jump rates depend on the next-nearest neighbor occupation.}
\label{fig:illustration}
\end{figure}

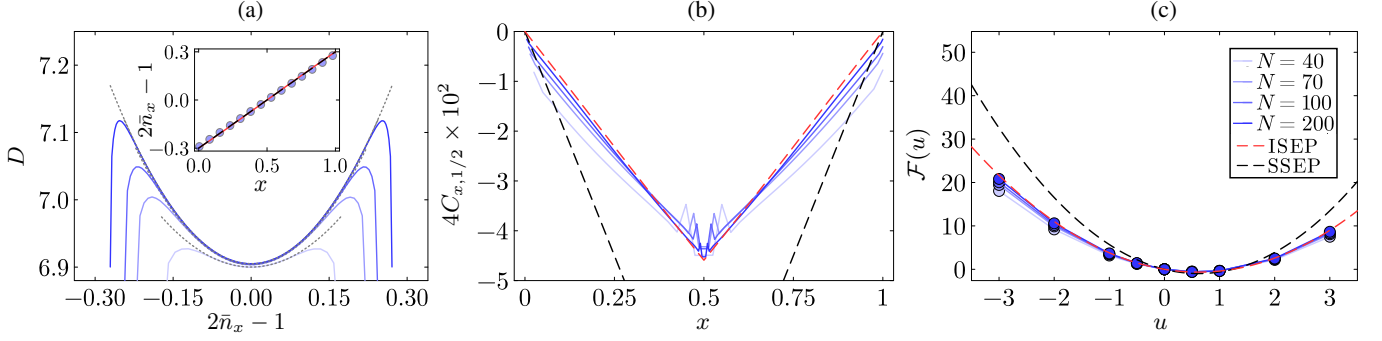
\begin{figure*}[!t]
\centering
\scalebox{0.958}[1]{
\begin{overpic}[width=0.32\textwidth]
    {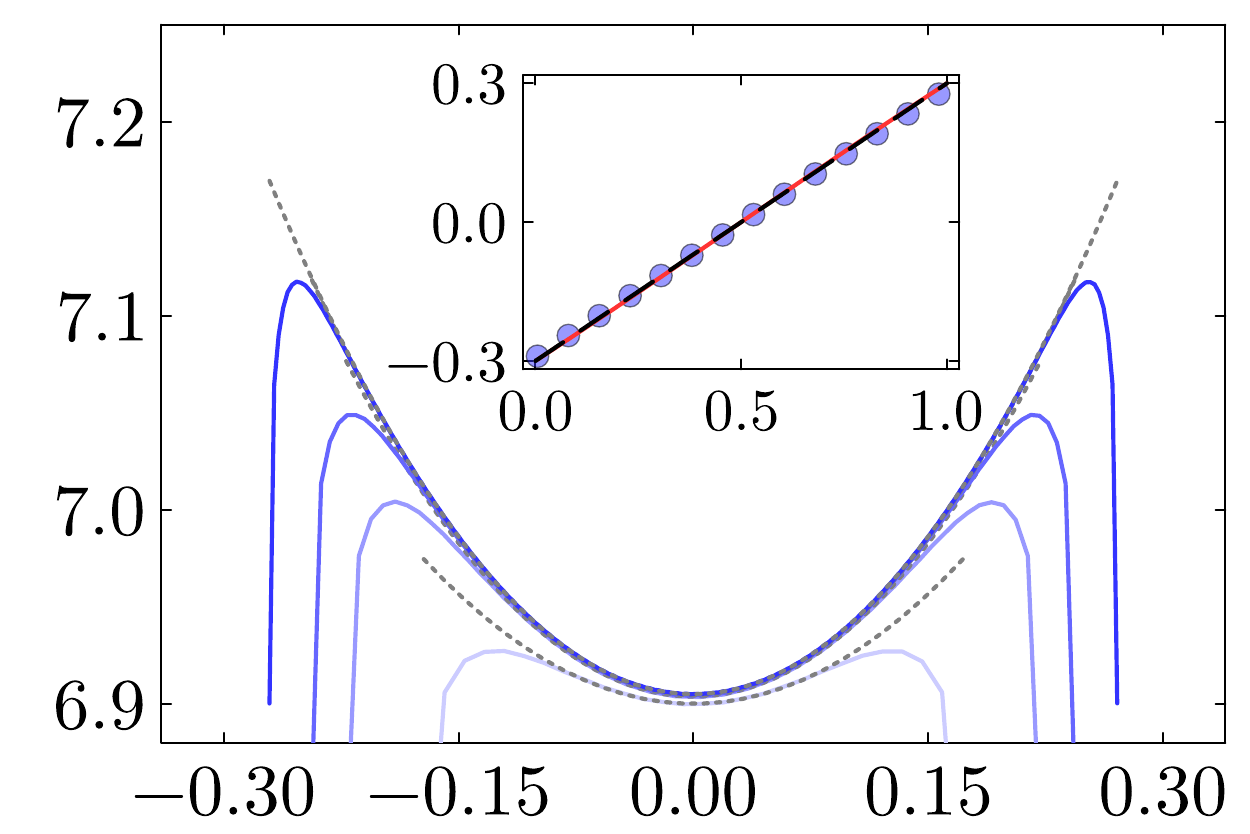}
    
    \put(52.3,68){(a)}
    \put(56.2,28){$x$}
    \put(28.4,47){\rotatebox[origin=c]{90}{ {\footnotesize$2\bar n_x -1$}}}

    \put(44.9,-4){$ 2\bar n_x -1$}
 \put(-3,34){\rotatebox[origin=c]{90}{$D$}}
\end{overpic}%
}
\hspace{+3.5mm}%
\begin{overpic}[width=0.32\textwidth]
    {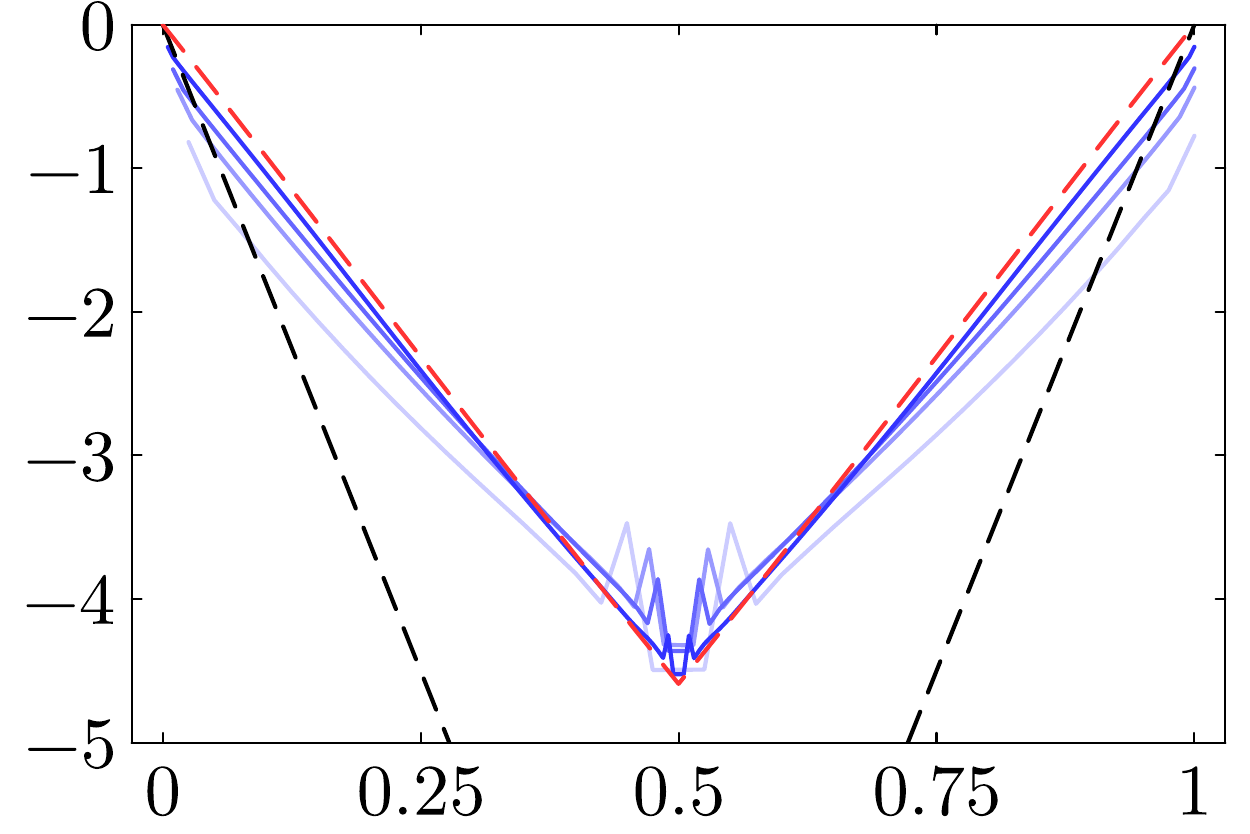}
    \put(50.5,68){(b)}
    \put(52.5,-4){$x$}
   \put(-7,35){\rotatebox[origin=c]{90}
        {$4 C_{x,1/2} \times 10^2$}}
\end{overpic}%
\hspace{+4mm}%
\begin{overpic}[width=0.32\textwidth]
{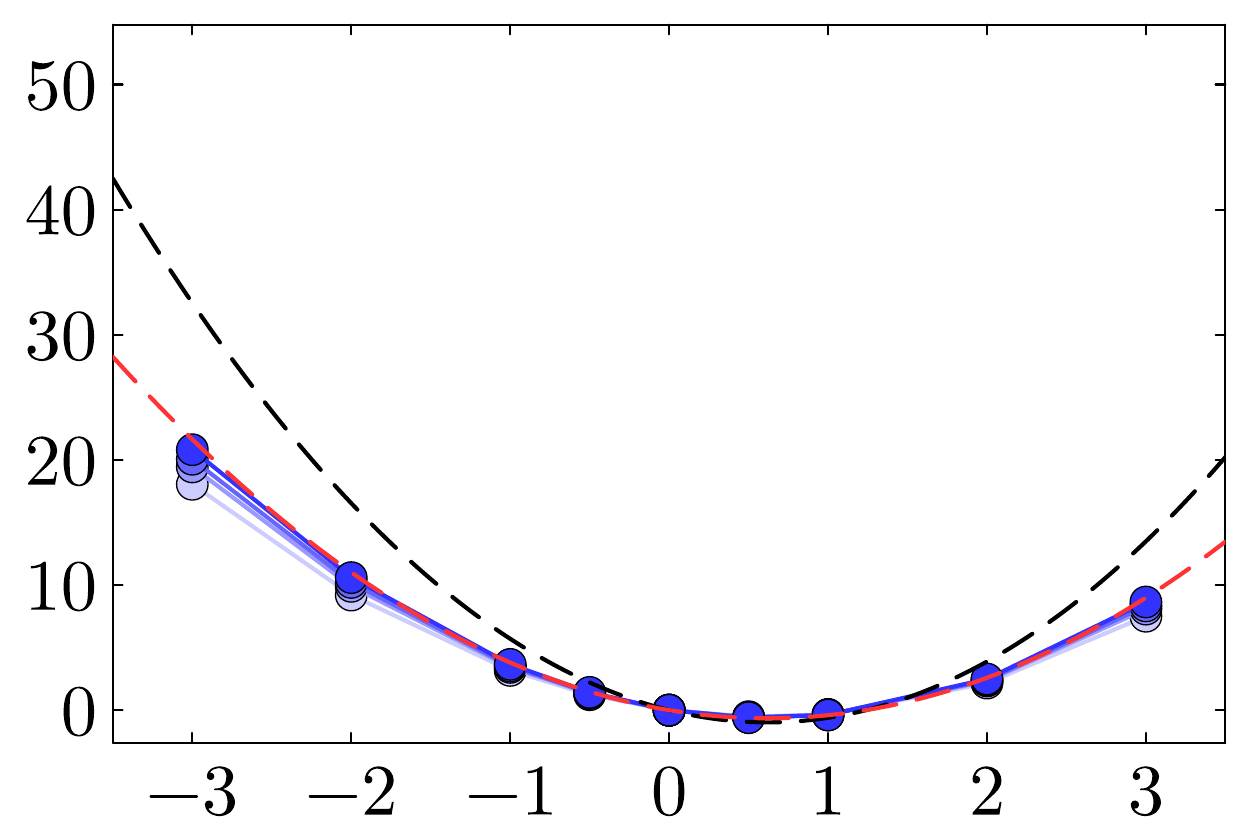}
    \put(68.5,30.8){\includegraphics[width=0.08\textwidth]{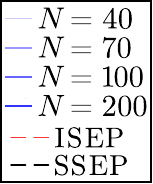}
    }
     \put(50.3,68){(c)}
    \put(51,-4){$u$}
    \put(-6,35){\rotatebox[origin=c]{90}
        {$\mathcal{F}(u)$}}
\end{overpic}
 
\caption{Comparison of tensor-network results for the microscopic model \eqref{eq:DynamicsAve_rho} and the ISEP description. 
(a) Local diffusivity reconstructed from the NESS magnetization profile and current, $D=-\bar J_x/\nabla\bar n_x$, plotted against the local magnetization $\langle\hat\sigma^{z}_{j=xN}\rangle=2\bar n_x-1$.  The diffusivity in the bulk is well converged with system size at the maximum length $N=200$ considered. Fitting (gray dotted) Eq.~\eqref{eq:D(n)} to the bulk data  fixes the two effective couplings $( D_0, \lambda)$. The inset plot demonstrates the comparison of the magnetization profile with its analytical prediction for ISEP (see SM). 
(b) Connected spin--spin correlation $N\langle\hat\sigma^{z}_{i=xN}\hat\sigma^{z}_{j=yN}\rangle^{c}=4C_{x,y}$ at fixed $y=1/2$ (regular part $C^{\rm neq}_{x,y}$; the contact term at $x=y$ is not included). (c) Rescaled cumulant generating function $\mathcal{F}(u)$ versus the counting field $u$.
Using the values $( D_0, \lambda)$ for the largest $N$ from (a), we obtain the ISEP--MFT (red) curves in (b) and (c), with which our numerics show very good agreement and convergence with $N$, while they deviate significantly from the non-interacting case of SSEP (black curves).
Tensor-network simulation for all panels is performed at
$\varepsilon=-1$, $\eta=0.2$, $\Delta=0.5$, $\gamma=1$, $\mu=0.3$, up to times $t_{\rm max}=3000$, for $N\in\{40,70,100,200\}$, bond dimension $\chi=200$ and TEBD2 algorithms with $\delta t =0.02$ time step. The legend inset of panel (c) applies to all the panels.}
\label{fig:ISEPnumerical} 
\end{figure*}

 In this Letter, we close this gap. Starting from the boundary-driven XXZ spin chain with bulk dephasing, see Fig.~\ref{fig:illustration}, we derive---for the first time and directly from the microscopic Lindblad dynamics---an emergent classical MFT characterized by density-dependent diffusivity and mobility. This microscopic-to-hydrodynamic connection allows us to obtain exact predictions for the system's stationary properties and current fluctuations, which we benchmark against tensor-network simulations of the Lindblad dynamics~\cite{AJDaley_2004,zwolak04}, finding excellent agreement; see Fig.~\ref{fig:ISEPnumerical}.

Beyond the XXZ model itself, our results demonstrate that the emerging universality of MFT extends beyond constant-diffusivity models to genuinely interacting quantum systems, with microscopic quantum mechanics entering only through the transport coefficients. Thus, our work opens the way to a systematic fluctuating hydrodynamic description of interacting diffusive quantum matter.
\medskip

\prlsection{Model and strategy}\label{sec:model}
We first introduce the microscopic model. We consider the one-dimensional XXZ spin chain,
\be\label{eq:Hxxz}
\hat H_\text{xxz} :=\sum_{j=1}^{N-1}\left[\varepsilon\left(\hat\sigma_{j}^{x}\hat\sigma_{j+1}^{x}+\hat\sigma_{j}^{y}\hat\sigma_{j+1}^{y}\right)+\Delta\hat\sigma_{j}^{z}\hat\sigma_{j+1}^{z}\right],
\ee
where $\hat\sigma_j^{a}$ ($a=x,y,z$) are standard Pauli operators acting on site $j$. The chain is subject to both bulk dephasing and boundary driving. For this model, the dynamics of the density matrix $\hat\rho$ is governed by the GKLS \cite{Lindblad1976,Gorini1976} equation
\begin{equation}
\label{eq:DynamicsAve_rho}
\partial_{t}\hat\rho=-i[\hat H_\text{xxz},\hat\rho]+{\cal L}_{\eta}\left(\hat\rho\right)+{\cal L}_\text{bdy}\left(\hat\rho\right).
\end{equation}
The non-unitary part of the dynamics is generated by
\begin{align}
\label{eg:LindDissipatorsMicro}
\!\!{\cal L}_{\eta}\left(\hat\rho\right) \! :=\eta\sum_{j=1}^{N}{\cal D}_{\hat \sigma_{j}^{z}}\left(\hat\rho\right);\; {\cal L}_\text{bdy}\left(\hat\rho\right)\! :=\!\!\!\!\sum_{\substack{p \in \left\{1,N\right\} \\
a\in\left\{ +,-\right\} }
}\!\!\!\!\Gamma_{p,a}{\cal D}_{\hat\sigma_{p}^{a}}\left(\hat\rho\right), 
\end{align}
where we introduce the Lindblad dissipators, ${\cal D}_{\hat L_j}\left(\hat\rho\right):=\hat L_{j}\hat\rho \hat L_{j}^{\dagger}-\frac{1}{2}\{ \hat L_{j}^{\dagger}\hat L_{j},\hat\rho\}$, with reservoir couplings $\Gamma_{1,a}=\gamma(1-a\mu)$, $\Gamma_{N,a}=\gamma(1+a\mu) $. Here, $\gamma$ controls the coupling strength to the baths, while $\mu$ implements the magnetization imbalance set by the reservoirs at the two boundaries. In particular, when $\mu\neq0$, the system is driven at long times towards a Non-Equilibrium Steady State (NESS) carrying a finite current. Importantly, the bulk terms conserve the total magnetization. 
An illustration of the microscopic setup is given in Fig.~\ref{fig:illustration}.

Due to the presence of interactions, the theory is not Gaussian and the equations of motion for the two-point function no longer close. Furthermore, the presence of bulk noise breaks the integrability of the XXZ chain, making the transport diffusive and preventing us from exploiting the analytical Matrix Product State (MPS) structure of NESS in the absence of noise \cite{ProsenExactXXZ,kps2013,XXZ_Exact_Clerk,Prosen_MPS_XYZ}. 

However, a renormalization group argument indicates that the system's large-scale behavior is governed by \textit{an effective strongly interacting diffusive regime} that admits a simpler description.
Indeed, fixing the scale of the free XX part of the Hamiltonian~\eqref{eq:Hxxz}, a tree-level renormalization analysis \cite{Kamenev2011} shows that the effective couplings at scale $\ell$ are given by~\cite{Cardy1996}
\begin{align}
\eta(\ell) & \sim\Delta(\ell)\sim e^{\left(2-d\right)\ell}, \label{RG_scaling}
\end{align}
where $d$ is the spatial dimension---see the Supplemental Material (\hyperlink{SM}{SM}) for a derivation. Both terms are therefore relevant in $d=1$. For $\Delta=0$, the flow towards strong $\eta$ was verified numerically in Ref.~\cite{Costa_emergence_universality_2026_PRL}. One thus expects the large-scale physics to be governed by a diffusive strong-coupling fixed point \footnote{The boundary terms also scale as $e^{\left(2-d\right)\ell}$ and therefore flow towards strong coupling. Consequently, the boundary magnetization is fixed to its equilibrium value at leading order in $\ell$.}. Expanding perturbatively in the inverse noise strength $\eta^{-1}$ and at fixed ratio $\Delta/\eta$, we derive below an effective model corresponding to an interacting version of the SSEP~\cite{iqsep-paper}. Although interactions still generate an infinite hierarchy of coupled correlations, we show that these correlations factorize at leading order in the system size, yielding a closed set of equations.

The effective couplings $\eta(\ell)$ and $\Delta(\ell)$ in the strong-coupling regime 
remain unknown, as determining them would require a non-perturbative solution of the RG flow. Computing such a solution is in general a formidable task in itself, but fortunately, it is not required for our purposes. Indeed, the MFT provides an effective description of fluctuations once the diffusivity and mobility are determined microscopically. These transport coefficients are linked by a fluctuation–dissipation relation \cite{mft2015rmp}, leaving only the diffusivity to be determined. We will show that the latter can be expressed analytically in terms of the effective parameters. These parameters can then be extracted numerically, making the theory predictive for all other single-replica quantities.

\medskip

\prlsection{Strong-coupling model}\label{sec:StrongCoupling}
We now work in the strong-coupling limit in $d=1$ and introduce rescaled notations for the effective couplings 
$\eta_{\rm eff}:=\eta(\ell){e^{-\ell}} $, $\Delta_{\rm eff} :=\Delta(\ell)e^{-\ell}$, and time $s:= te^{-\ell}$. In this limit, the anisotropy and dephasing terms exponentially suppress the off-diagonal components of the density matrix in the product-state basis $|\mathbf{n}\rangle:=|n_1,\ldots,n_N\rangle$, with $\hat n_j|\mathbf{n}\rangle=n_j|\mathbf{n}\rangle$, $n_j\in\{0,1\}$, and $\hat n_{j}:=(\hat{\mathbb{I}}+\hat\sigma_{j}^{z})/2$. The resulting classical slow-mode manifold consists of diagonal states~\cite{BernardJinShpielberg_2018}, $\bar{\rho}_s:=\mathcal{P}(\hat\rho_s)=\sum_{\mathbf{n}} \Pi_s(\mathbf{n}) |\mathbf{n}\rangle\langle\mathbf{n}|$, where $\mathcal{P}$ is the corresponding projector and $\Pi_s(\mathbf{n})\in[0,1]$ is the probability of the spin configuration $\mathbf{n}$. On long timescales $t\sim e^\ell$, the dynamics restricted to this manifold is governed by, to leading order in $e^{-\ell}$,~\cite{Schrieffer_Wolff_Kessler2012,PhysRevA.93.022312,BauerBernardJin_Stoqdissipative,Strong_dissipation_Popkov2021} 
\begin{equation}
\label{eq:strongcouplingMaster}
\partial_{s}\bar{\rho}=\mathcal{P}{\cal L}\left({\cal L}_{b}^{\perp}\right)^{-1}{\cal L}\mathcal{P}\left(\bar{\rho}\right)
\end{equation}
where ${\cal L}_b(\bullet):=-i[\sum_{j}\Delta\hat\sigma_{j}^{z}\hat\sigma_{j+1}^{z},\bullet]+{\cal L}_{\eta}(\bullet)$, and ${\cal L}(\bullet)=-i[\sum_{j}\varepsilon\left(\hat\sigma_{j}^{x}\hat\sigma_{j+1}^{x}+\hat\sigma_{j}^{y}\hat\sigma_{j+1}^{y}\right),\bullet]$. The slow-mode manifold is the kernel of ${\cal L}_{b}$, while ${\cal L}_{b}^{\perp}$ denotes its restriction to the complementary subspace. Although Eq.~\eqref{eq:strongcouplingMaster} governs only the projected state $\bar{\rho}$, the operator ${\cal L}$ generates a quantum jump out of the slow-mode manifold, which must be compensated by a second jump back to contribute to the effective dynamics.

Explicitly, the effective dynamics~\eqref{eq:strongcouplingMaster}  decomposes into four jump processes,
\begin{equation}\label{eq:eff-dyn}
\partial_{s}\bar\rho=\sum_{j}\!\sum_{\alpha,\beta\in\left\{ \uparrow,\downarrow\right\}}\!\left[{\cal D}_{\hat L^{\alpha,\beta}_{j;+}}(\bar\rho)+{\cal D}_{\hat L^{\alpha,\beta}_{j;-}}(\bar\rho)\right] +{\cal L}_\text{bdy}(\bar\rho),
\end{equation}
with jump operators 
$\hat L^{\alpha,\beta}_{j;\pm} \!=\!\sqrt{A_{\alpha,\beta}}\,\hat{\mathbb{P}}_{j-1}^{\alpha}\hat\sigma_{j}^{\pm}\hat\sigma_{j+1}^{\mp}\hat{\mathbb{P}}_{j+2}^{\beta}$,
where $\hat{\mathbb{P}}^{\uparrow,\downarrow}:=\frac{\hat{\mathbb{I}}\pm\hat\sigma^{z}}{2}$
and
\begin{align}\label{eq:A-ampl}
A_{\uparrow,\uparrow} & =A_{\downarrow,\downarrow}=\frac{2\varepsilon^{2}}{\eta_{\text{eff}}},\  A_{\uparrow,\downarrow}=A_{\downarrow,\uparrow}=\frac{2\varepsilon^{2}/\eta_\text{eff}}{1+\left(\Delta_{\text{eff}}/\eta_{\text{eff}}\right)^2},
\end{align}
see \hyperlink{SM}{SM} for details of the derivation. The invariance of these amplitudes under spin flip is a consequence of the underlying $\mathbb{Z}_2$ symmetry of the effective dynamics. Moreover, as anticipated, they depend on the effective values $\eta_\text{eff}$ and $\Delta_\text{eff}$.
The Heisenberg equation of motion for the density obtained from Eq.~\eqref{eq:eff-dyn} is a conservation law, $\partial_s \hat{n}_j = \hat{J}_{j-1}-\hat{J}_{j}$. The current is $\hat{J}_j:=-\hat{D}_j(\hat{n}_{j+1}-\hat{n}_j)$, with density-dependent diffusivity operator
\be\label{eq:interaction}
\hat D_j:=D_0\big[1+ \lambda \hat n_{j-1}\big(1-\hat n_{j+2}\big)+ \lambda \hat n_{j+2}\big( 1- \hat n_{j-1}\big)\big]
\ee
and with $D_0=A_{\downarrow\downarrow}$, $\lambda=\big[A_{\uparrow\downarrow}-A_{\downarrow\downarrow}\big]/A_{\downarrow\downarrow}$, see \hyperlink{SM}{SM} for details of the derivation. Notice that the particle-hole symmetry of the model is explicitly enforced.

\medskip

\prlsection{ISEP and its MFT description}
The time evolution of the probability distribution $\Pi_s(\mathbf{n})$, obtained from Eq.~(\ref{eq:eff-dyn}), encodes a Markov process whose transition rules are depicted in Fig.~\ref{fig:illustration}. It corresponds to an exclusion process in which particles hop between nearest-neighbor sites at rates conditioned on the occupation of the next-to-nearest-neighbor sites.
It thus differs from the simple exclusion process, which would correspond to the non-interacting XX model ($\Delta=0$), via a dressing of the local hopping rate by the factor $\langle \mathbf n|\hat D_j|\mathbf n\rangle$;  see \hyperlink{EM}{End Matter}.
In a parallel work~\cite{iqsep-paper}, the same type of process emerged in the single-replica description of the Interacting Quantum Symmetric Exclusion Process, dubbed IQSEP. For this reason, we shall refer to the present Markov chain as ISEP, as the single-replica only corresponds to the classical part \cite{BernardJin_QSSEP}.

Similar Markov processes have been considered previously, notably facilitated exclusion processes and related models~\cite{Rossi2000,deOliveira2005,Barraquand2025,Kob1993,Gabel2010,Baik2018,Goldstein2019,Ayyer2023}, for which the emergence of nonlinear hydrodynamics at large scales has been investigated~\cite{Blondel2020,DaCunha2026,Gonalves2009,Funaki1991}. We show below that a similar hydrodynamic description controls the spin correlations of the dephased XXZ model at large scales.

Following this program, we introduce the fluctuating hydrodynamic equations of MFT. We take the continuum limit $N\to\infty$, with rescaled position $x:=j/N$ and diffusive time $\tau:=s/N^2$. The MFT density and current fields obey~\cite{SadhuDerrida2016}
\begin{align}
\partial_\tau \n_x &=-\nabla \jj_x,\nonumber\\
\jj_x &=-D\left(\n_x\right)\nabla \n_x
+\sqrt{\sigma(\n_x)/N}\,\xi_x,
\label{eq:fluctuating-hydrodynamics}
\end{align}
where $\xi_x(\tau)$ is Gaussian white noise with zero mean and variance $\mathbb{E}[\xi_x(\tau)\xi_y(\tau')]
=\delta(x-y)\delta(\tau-\tau')$. Here, $\mathbb{E}[\bullet]$ denotes the average over the noise. Consistently with the $N^{-1/2}$ noise amplitude in Eq.~\eqref{eq:fluctuating-hydrodynamics}, we parametrize the MFT density field as 
\be
\n_x=\bar n_x+\delta\n_x/\sqrt{N},
\ee
where $\bar n_x:=\mathbb{E}[\n_x]$ is a non-fluctuating density profile, and $\mathbb{E}[\delta\n_x]=0$.

Our aim is to show that the coarse-grained dynamics of the ISEP reproduces this structure at leading order in $1/N$, and to determine the transport coefficients $D(n)$ and $\sigma(n)$ from the microscopic dynamics. At this stage, Eq.~\eqref{eq:fluctuating-hydrodynamics} is the macroscopic form that we seek to recover, rather than an additional assumption about the ISEP. To connect MFT to the ISEP, we identify the density profiles $\bar n_x=\left.\langle\hat n_j\rangle\right|_{x=j/N}$, with $\langle\bullet\rangle:={\rm tr}(\bar\rho_s\,\bullet)$. This quantity evolves according to the Heisenberg equation associated with Eq.~\eqref{eq:eff-dyn}. Because of interactions, however, this equation does not close, unlike in the (Q)SSEP, as Eq.~\eqref{eq:interaction} illustrates (see also \hyperlink{EM}{End Matter}).\\
\indent
The key assumption that allows us to make progress is that, in the large-$N$ limit, density correlations obey the same scaling underlying MFT~\cite{mft2015rmp}, namely
\begin{equation}
\left\langle
\hat n_{i_1}\cdots\hat n_{i_m}
\right\rangle^c
={\cal O}\left(N^{1-m}\right).
\label{eq:MFT_scaling_density}
\end{equation}
Here, the sites $i_k$ are all distinct, and $\langle\bullet\rangle^c$ denotes the $m$-point cumulant of density operators inside brackets. Such an assumption encodes the standard statistical physics fact that correlations become weak as the system size grows, both in and out of equilibrium, provided the system is away from criticality. Note that Eq.~\eqref{eq:MFT_scaling_density} is consistent with the weak-noise expansion of Eq.~\eqref{eq:fluctuating-hydrodynamics} which then gives the same $N^{1-m}$ scaling at distinct macroscopic positions. 

The consequence used below is that density operators with disjoint supports factorize at leading order in $1/N$. If $\hat O_{\hat n}:=\hat n_{i_1}\cdots\hat n_{i_m}$ and $\hat O_{\hat n}'$ have disjoint supports, then
\be\label{eq:factorization}
\langle\hat O_{\hat n}\hat O_{\hat n}'\rangle =\langle\hat O_{\hat n}\rangle\langle\hat O_{\hat n}'\rangle+{\cal O}(N^{-1}).
\ee
Using Eq.~\eqref{eq:factorization}, the Heisenberg equations therefore close at leading order and yield the diffusive hydrodynamics derived below. We verify Eq.~\eqref{eq:MFT_scaling_density} numerically for the second- and third-order cumulants of the microscopic model~\eqref{eq:DynamicsAve_rho}; see the \hyperlink{EM}{End Matter}.

As shown below, the factorization property~\eqref{eq:factorization} allows us to recover the MFT structure and, in turn, to fix $D(n)$ and $\sigma(n)$ from the microscopic dynamics. We first compute the diffusivity $D(n)$. The microscopic conservation law $\partial_s\hat n_j=\hat J_{j-1}-\hat J_j$ yields, after coarse graining,
\begin{equation}\label{eq:conservation}
\partial_\tau \bar n_x=-\nabla \bar J_x ,
\end{equation}
with averaged rescaled current $\bar J_x=-N\langle \hat J_j\rangle|_{x=j/N}$. Using Eq.~\eqref{eq:factorization}, this current asymptotically fulfills Fick's law,
\begin{equation}\label{eq:Fick}
\bar J_x=-D(\bar n_x)\nabla\bar n_x\left[1+{\cal O}(N^{-1})\right],
\end{equation}
with density-dependent diffusivity
\begin{equation}\label{eq:D(n)}
D(\bar n)=D_0\left[1+2\lambda\bar n(1-\bar n)\right].
\end{equation}

We now turn to the mobility $\sigma(n)$, which can be fixed by a simple equilibrium argument. Indeed, at equilibrium ($\mu=0$), the two-point cumulant takes the simple form
\be
\left\langle \hat n_i\hat n_j\right\rangle^c=\bar n(1-\bar n)\delta_{ij},
\ee
and gives, at large scales, the compressibility $\chi(\bar n)=\bar n(1-\bar n)$. The Einstein relation~\cite{mft2015rmp} then fixes the mobility via
\be\label{eq:mobility}
\frac{\sigma(\bar n)}{D(\bar n)}=2\bar n(1-\bar n).
\ee
Out of equilibrium, the same expression for $\sigma(n)$ can be read directly from the equation of motion for the second density cumulant, as discussed below and derived in the \hyperlink{SM}{SM}.

Thus, our only remaining task is to determine numerically the values of the coefficient $D_0$ and $\lambda$ appearing in \eqref{eq:D(n)}. In order to do so, we perform tensor-network simulation of the microscopic model \eqref{eq:DynamicsAve_rho}, where we time-evolve the vectorized density matrix MPS representation towards the steady state $\hat\rho_{\infty}$ \cite{zwolak04,Jaschke_2019}. The numerical results confirm the density-dependent diffusivity from Eq.~\eqref{eq:D(n)}. Fig.~\ref{fig:ISEPnumerical}(a) shows the numerically obtained dependence on the local magnetization expectation value, with the diffusivity $D$ extracted from Fick's law. Here, the average magnetization and magnetization gradient are fitted within a window of 12 sites around site $i$, and $\bar{J} = {\rm tr}(\hat{J}_i \,\hat\rho_{\infty})$. We discard boundary layers of width $N/5$ at both edges to avoid contact-resistance effects and recover the bulk transport.  Observed convergence with respect to system sizes $N \in \{40,70, 100, 200\}$ implies that fitting $D_0$ and $\lambda$ from $N=200$ results yields nearly thermodynamic values, see \hyperlink{SM}{SM}.

\medskip

\prlsection{
Macroscopic profile and fluctuations
} Once $D$ and $\sigma$ are fixed, the ISEP/MFT description becomes predictive for single-replica density observables. Combining Eqs.~\eqref{eq:conservation} and~\eqref{eq:Fick} yields the nonlinear diffusion equation $\partial_\tau\bar n_x=\nabla\!\left[D(\bar n_x)\nabla\bar n_x\right]+{\cal O}(N^{-1})$, with boundary densities $\bar n_{x\in\{0,1\}}=(1\pm\mu)/2$. Its stationary solution gives the density profile shown in the inset of Fig.~\ref{fig:ISEPnumerical}(a).

We next consider the equal-time connected two-point correlation $C_{x,y}:=\lim_{N\to\infty}N\langle\hat n_i\hat n_j\rangle^c|_{x=i/N;\,y=j/N}$. As shown in the \hyperlink{SM}{SM}, its microscopic equation of motion yields
\begin{equation}\label{eq:eq-for-C}
\left(\partial_{\tau}-\Delta_{x}^{D}-\Delta_{y}^{D}\right)C_{x,y} = \partial_{x}\partial_{y}\left[\sigma\left(\bar{n}_{x}\right)\delta\left(x-y\right)\right],
\end{equation}
where $\Delta_{x}^{D}\bullet:=\partial_{x}^{2}\!\left[D\left(\bar{n}_{x}\right)\bullet\right]$ and boundaries $C_{0,y}=C_{1,y}=C_{x,0}=C_{x,1}=0$.
The same equation follows within MFT by identifying $C_{x,y}=\mathbb{E}[\delta\n_x\delta\n_y]$, by linearizing Eq.~\eqref{eq:fluctuating-hydrodynamics} around $\bar n_x$, and averaging over the noise. This provides a direct consistency check of the MFT description. In the stationary state, its numerical solution agrees with the NESS spin correlation $\langle\hat\sigma_{i=xN}^z\hat\sigma_{j=yN}^z\rangle^c =4C_{x,y}/N+{\cal O}(N^{-2})$ obtained from tensor-network simulations; see Fig.~\ref{fig:ISEPnumerical}(b).

As a final test of MFT, we consider the full-counting statistics of the integrated current. Let $Q_s$ denote the net charge transferred through one of the boundaries during the time interval $[0,s]$. Its moment generating function,
\begin{equation}\label{eq:MomGen}
Z_s(u):=\mathbb{E}\left[e^{uQ_s}\right] \underset{s\to\infty}{\asymp} \exp\left[\tau N {\cal F}(u)\right],
\end{equation}
defines the rescaled cumulant generating function ${\cal F}(u)$ in the diffusive long-time limit $s=\tau N^2$.
The calculation of ${\cal F}(u)$ within MFT is standard and has been detailed, for instance, in Refs.~\cite{BodineauDerridaAP_2004,Lecomte_2010,Bodineau2010,Derrida2011,Derrida_2025}. It assumes that the saddle-point profile minimizing the MFT action becomes stationary at long times, an assumption that is generally well justified \footnote{Indeed, the breakdown of this assumption is intimately connected to the emergence of dynamical phase transitions (DPTs). A sufficient condition ensuring stationarity and excluding DPTs is $D'(\bar{n})\sigma'(\bar{n})\geq D(\bar{n}) \sigma''(\bar{n})$ for any $\bar{n}$~\cite{Bertini_2006}. One can check that ISEP satisfies this assumption for any value of its parameters}. We report the derivation in the \hyperlink{SM}{SM} and compare the resulting prediction with numerical simulations in Fig.~\ref{fig:ISEPnumerical}(c), again finding excellent agreement.

\medskip

\prlsection{Conclusion} We have established a MFT description of the non-equilibrium steady state for the boundary-driven XXZ spin chain with bulk dephasing. Starting from the microscopic Lindblad dynamics, we derived in the strong-coupling scaling limit an effective interacting exclusion process, the ISEP, governing the slow degrees of freedom. Coarse-graining this stochastic dynamics yields a nonlinear diffusion equation with a density-dependent diffusivity and mobility, from which we obtained the stationary density profile, long-range density correlations, and FCS of the current. All predictions are in excellent agreement with tensor-network simulations of the microscopic quantum dynamics.

To the best of our knowledge, together with the parallel work~\cite{iqsep-paper}, this provides the first explicit derivation of interacting hydrodynamics, characterized by density-dependent transport coefficients, directly from a microscopic noisy quantum many-body model. This demonstrates that the emerging universality of classical MFT extends beyond the constant-diffusivity class of the symmetric simple exclusion process to genuinely interacting quantum matter, with microscopic quantum mechanics entering only through the transport coefficients $D(\bar n)$ and $\sigma(\bar n)$. On a practical level, the closure of the BBGKY-like hierarchy \cite{Bogoliubov1946,BornGreen1946,Kirkwood1946}, enabled by MFT scaling, provides a powerful and broadly applicable framework for the systematic study of interacting quantum diffusive systems. Such a closure technique was also used in~\cite{iqsep-paper}.


Several directions naturally follow. A first is to delineate the universality of this construction and determine how broadly effective interacting exclusion processes emerge from noisy quantum many-body dynamics. In particular, it remains open whether density-dependent transport coefficients can drive the critical phenomena familiar from classical MFT, such as dynamical phase transitions, in other noisy quantum systems. A second is to characterize the renormalization flow relating the microscopic couplings to the effective transport coefficients of the hydrodynamic theory. Finally, perhaps the most intriguing challenge is to move beyond the classical MFT description and systematically capture the genuinely quantum corrections to large deviations, which are expected to become visible only beyond the leading hydrodynamic order; identifying the appropriate interacting quantum extension of MFT~\cite{Bernard2021} remains an important open problem. As a first step in this direction, the parallel work~\cite{iqsep-paper} identified the structure of multi-replica coherence loops in the presence of interactions. These results provide a natural starting point to explore quantum corrections to MFT in future work.

\medskip
\prlsection{Acknowledgements.} 
SS and DB acknowledge F. H\"ubner for collaboration on related topics. AC acknowledges M. Coppola for useful discussions. SS is supported by the MSCA Grant No.~101103348 (GENESYS). DB is partly supported by the CNRS, the ENS and the Simons Foundation via the Simons Collaboration on Probabilistic Paths to QFT. TJ is funded by the ANR-25-CE57-2088 JCJC (QuDi). AC and ZL acknowledge support from ERC StG 2022 project DrumS by Horizon Europe, Grant Agreement 101077265, and the P1-0044 program of the Slovenian Research and Innovation Agency (ARIS). JC is supported by Fundação para a Ciência e Tecnologia (FCT) through grants No.2022.11940.BD and UID/PRR2/04540/2025 (\href{https://doi.org/10.54499/UID/PRR2/04540/2025}{DOI}) to the I\&D unit Centro de Física e Engenharia de Materiais Avançados (CeFEMA). JDN is funded by ERC StG No.~101042293 (HEPIQ) and ANR-22-CPJ1-0021-01.
Views and opinions expressed are those of the authors only and do not necessarily reflect those of the European Union or the European Research Council Executive Agency. Neither the European Union nor the granting authority can be held responsible for them. 

\bibliography{bibliography.bib,biblio_tony.bib,ref_intro.bib} 

\begin{center}
\hypertarget{EM}{\Large \textbf{End Matter}}
\end{center}
\newcounter{emsection}
\newcommand{\emsection}[1]{%
  \refstepcounter{emsection}%
  \section*{\Roman{emsection}. #1}%
}
\begin{figure*}
    \centering
\scalebox{0.958}[1]{
\begin{overpic}[width=0.30\textwidth]
    {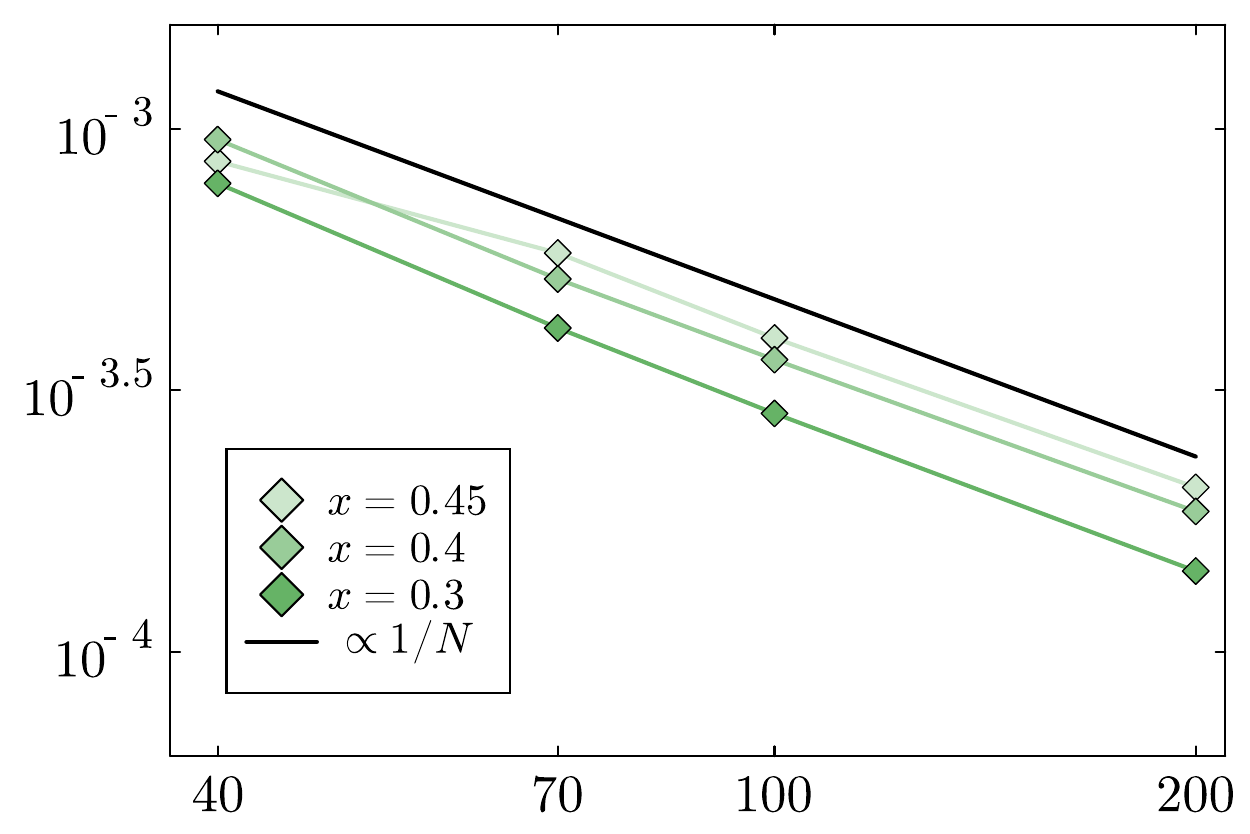}
    
    \put(50.3,58){(a)}
    \put(50.3 ,-4){$N$}
 \put(-6,35){\rotatebox[origin=c]{90}{$|\langle \hat\sigma_{i=xN}^z \hat \sigma^z_{j=N/2} \rangle^c |$}}
\end{overpic}%
}
\hspace{+4mm}%
\begin{overpic}[width=0.30\textwidth]
    {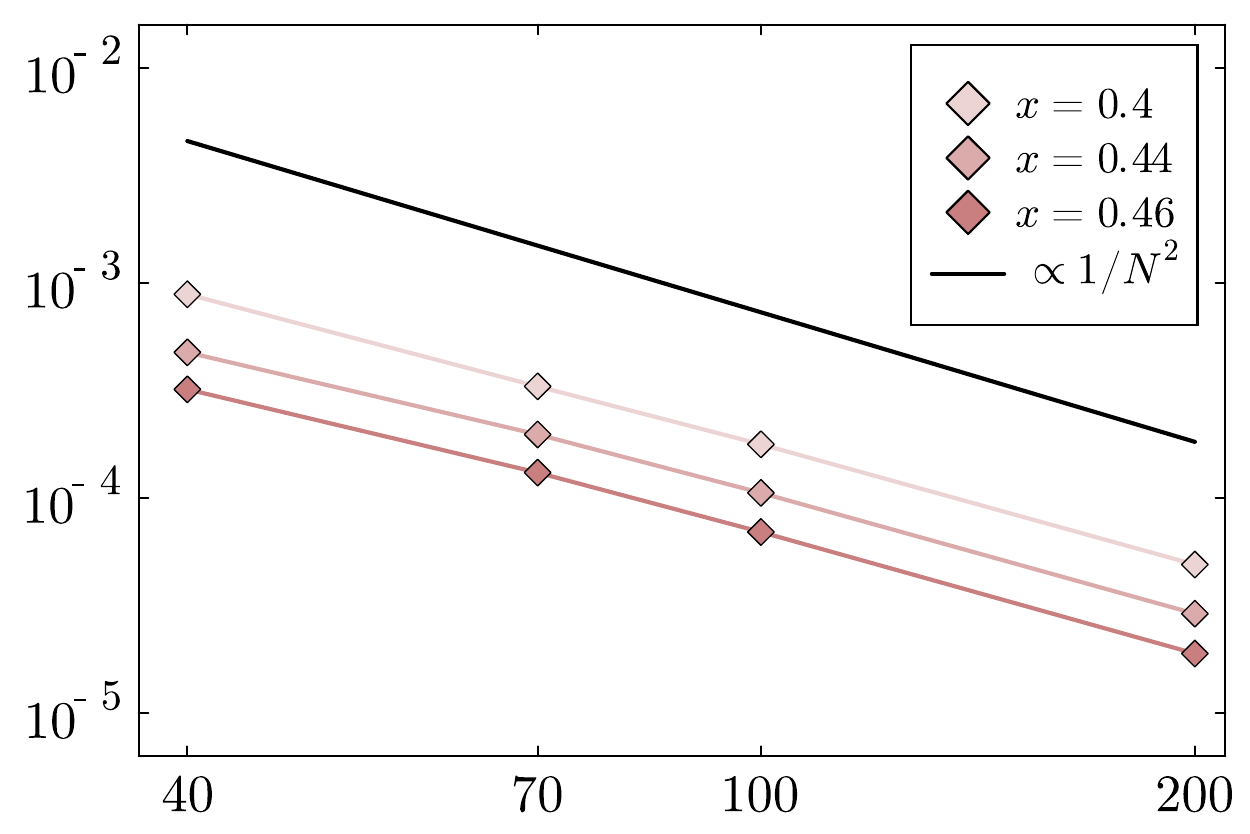}
    \put(50.3,58){(b)}
    \put(50.3,-4){$N$}
   \put(-6,34){\rotatebox[origin=c]{90}
        {$|\langle\hat \sigma^z_{N/2} \hat \sigma^z_{N/6} \hat \sigma^z_{i=xN} \rangle^c|$}}
\end{overpic}%
\hspace{+4mm}%
\begin{overpic}[width=0.30\textwidth]
    {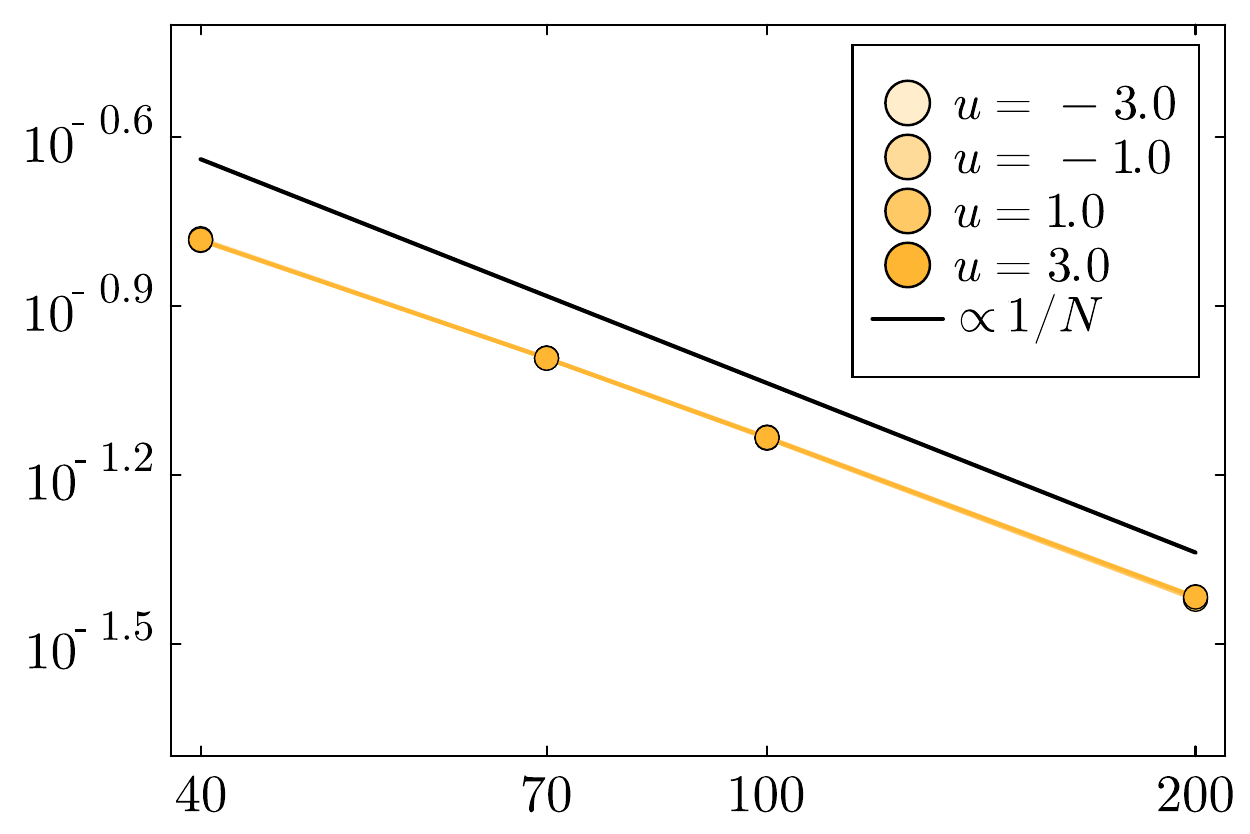}
     \put(50.3,58){(c)}
    \put(50.3,-4){$N$}
    \put(-6,35){\rotatebox[origin=c]{90}
        {$|1- N \Psi_N/\mathcal{F}|$}}
\end{overpic}

\caption{Finite-size scaling tests of the MFT
  description. (a) Connected spin correlator of the magnetization,
  $|\langle\hat\sigma^z_{i=xN}\hat\sigma^z_{j=N/2}\rangle^c|$, for
  $x=0.3,\,0.4,\,0.45$, which shows good agreement with the $1/N$ scaling predicted in Eq.~\eqref{eq:eq-for-C}. (b) Connected spin correlator $|\langle\hat \sigma^z_{N/2} \hat \sigma^z_{N/6} \hat \sigma^z_{i=xN} \rangle^c|$, with two sites  fixed and the third at
  $x=0.4,\,0.44,\,0.46$. The correlator shows good  agreement with the decay as $1/N^{2}$ predicted i.e.,\ the $m=3$ case of
  Eq.~\eqref{eq:MFT_scaling_density}. Panels (a) and (b) together
  validate the scaling used to extract our analytical results of  ISEP-MFT. (c) Relative deviation of the finite-size
  cumulant generating function $N\Psi_N(u)$ from the
  MFT prediction $\mathcal{F}(u)$, for $u=\pm1,\pm3$. We see that the numerical results demonstrate an approximate $1/N$ decay for all $u$, consistent with the $O(N^{-1})$ corrections at leading hydrodynamic order.}
\label{fig:MFT_N_scaling}
\end{figure*}
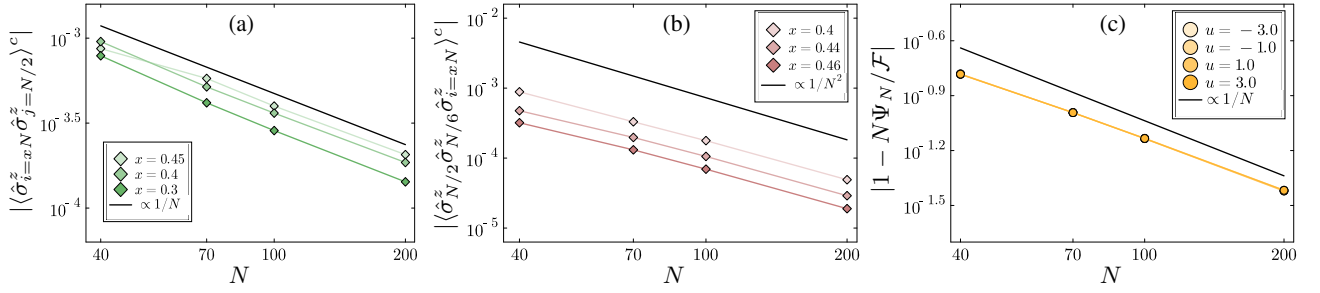  
\emsection{Additional details on the ISEP}

We report here the Markov process for the configuration probabilities $\Pi_s(\mathbf n)$ associated with the effective dynamics~\eqref{eq:eff-dyn}, leaving the details of the derivation to the \hyperlink{SM}{SM}. The ISEP master equation reads
\begin{align}
\partial_s \Pi_s(\mathbf n) = \sum_{j=1}^{N-2}\Big[&\underset{j\to j+1}{f(\mathbf n)}\Pi_s(\mathbf n^{j\to j+1})+\underset{j+1\to j}{f(\mathbf n)}\Pi_s(\mathbf n^{j+1\to j})
\nonumber\\
&-\big(\underset{j\to j+1}{f(\mathbf n)}+\underset{j+1\to j}{f(\mathbf n)}\big)\Pi_s(\mathbf n)\Big],
\end{align}
up to boundary terms. Here, $\mathbf n^{j\to k}$ denotes the configuration obtained by moving a particle from site $j$ to $k$, and the jumping rates between configurations are given by
\be\label{eq:transition-rates}
\underset{j\to k}{f(\mathbf n)}=\langle\mathbf n|\hat D_j|\mathbf n\rangle\,n_j(1-n_k).
\ee
The diffusivity operator $\hat D_j$ is diagonal in the configuration basis $|\mathbf n\rangle$ and it is given explicitly in Eq.~(\ref{eq:interaction}).
Interactions therefore amount to a configuration-dependent dressing of the local SSEP hopping rate $n_j(1-n_{j+1})$ across the link $j\to j+1$, which is now conditioned on the occupations of the two outer sites $j-1,j+2$ via $\hat D_j$. This process is illustrated in Fig.~\ref{fig:illustration}.

The master equation above explicitly makes the classical Markov process encoded by Eq.~\eqref{eq:eff-dyn}. For the hydrodynamic analysis, however, it is more convenient to work with the effective generator~\eqref{eq:eff-dyn} in the Heisenberg picture. For an arbitrary operator $\hat O$,
\be
\partial_s \hat O=\sum_j\sum_{\alpha,\beta}\left[{\cal D}^{*}_{\hat L^{\alpha,\beta}_{j;+}}(\hat O)+{\cal D}^{*}_{\hat L^{\alpha,\beta}_{j;-}}(\hat O)\right]+{\cal L}_{\rm bdy}^{*}(\hat O),
\ee
where ${\cal D}_{\hat L}^{*}$ denotes the adjoint Lindblad dissipator. For density observables $\hat O_{\hat n}=\hat n_{i_1}\dots\hat n_{i_m}$, the projectors entering the jump operators commute with both $\hat O_{\hat n}$ and the hopping operators; see the \hyperlink{SM}{SM} for further details. The equation of motion therefore reduces to
\be\label{eq:EM-Heisenberg}
\partial_s\langle\hat O_{\hat n}\rangle=\sum_j\left\langle\hat D_j\,{\cal L}^\text{ssep}_j(\hat O_{\hat n})\right\rangle+\left\langle{\cal L}_{\rm bdy}^{*}(\hat O_{\hat n})\right\rangle,
\ee
where
\be
{\cal L}^\text{ssep}_j(\bullet) :={\cal D}^{*}_{\hat\ell_j}(\bullet)+{\cal D}^{*}_{\hat\ell_j^\dagger}(\bullet),
\quad
\hat\ell_j=\hat\sigma_j^-\hat\sigma_{j+1}^+ .
\ee
Here, ${\cal L}^\text{ssep}_j$ is the local generator of the SSEP, describing symmetric particle hopping across a link with unit rate. Interactions enter Eq.~\eqref{eq:EM-Heisenberg} through the density-dependent operator $\hat D_j$. Since $\hat D_j$ itself contains density operators, the equation of motion above generates an infinite hierarchy. However, as explained in the main text, the MFT scaling in Eq.~\eqref{eq:MFT_scaling_density} allows this hierarchy to close at leading order in $1/N$, and thus to analytically investigate the correlation dynamics.
\emsection{Microscopic test of correlation scaling and current large deviations}
Here, we give more details on benchmarking the effective ISEP and the MFT predictions of
the main text against tensor-network simulations of the Lindblad dynamics, Eq.~\eqref{eq:DynamicsAve_rho}, respectively. In our implementation, the vectorized density matrix $\hat\rho$, is represented as an MPS and is evolved to the NESS using TEBD2. The vectorized dynamics has a weak $U(1)$ symmetry \cite{Buča_2012} associated with the conserved charge of the particle-number difference between the ket and bra copies; see \hyperlink{SM}{SM} for details. We exploit this symmetry through a
quantum-number-conserving MPS~\cite{SciPostPhysCodeb.4,Bernier2018LightCone},
which reduces the computational cost. Throughout our numerical
investigation, we fix the parameters of our model to the values mentioned in Fig.~\ref{fig:ISEPnumerical}.
For computational efficiency, and because we are interested in late times, we first evolve the system with bond dimension $40$ up to time of $t_{\rm max}/2$, and then increase it to $200$, to obtain a better estimation of NESS. The convergence in $\chi$ and $\delta t$ is discussed in the \hyperlink{SM}{SM}.


\emph{MFT scaling for correlations:}
The MFT scaling hypothesis~\eqref{eq:MFT_scaling_density}, which is crucial to our theoretical approach, is further supported by the 
$1/N$ and $1/N^2$ scaling of the two-point and three-point connected correlations, respectively, as we demonstrate in Figs.~\ref{fig:MFT_N_scaling}(a) and~(b).

\emph{Current large deviations:}
Finally, we test the current FCS predicted by our theory; see \hyperlink{SM}{SM}. The moment generating function $Z_s(u)$ of the integrated current defined in Eq.~\eqref{eq:MomGen} can be related to a tilted density matrix $\hat{\rho}_u$ via $Z_s(u)={\rm tr}(\hat{\rho}_u)$. The time evolution of $\hat{\rho}_u$ is generated by a tilted Liouvillian \cite{LebowitzSpohn1999_tilting_current,Esposito2009_tilting_current,Costa_emergence_universality_2026_PRL},
\begin{equation}
\label{eq:DynamicsAve_rhotilted}
\partial_t\hat\rho_u=-i\left(\hat H_\text{xxz}^{(u)}\,\hat\rho_u-\hat\rho_u \left(H_\text{xxz}^{(u)}\right)^{\dagger}\right)+{\cal L}_{\eta}\left(\hat\rho_u\right)+{\cal L}_\text{bdy}\left(\hat\rho_u\right)
\end{equation}
where we distributed the
counting field $u$ uniformly over the $N-1$ bonds of $\hat H_{\rm xxz}$ in Eq.~\eqref{eq:DynamicsAve_rho}.
The choice of $u$ is not necessarily uniform \cite{Costa_emergence_universality_2026_PRL,albert2026universalclassicalquantumfluctuations}, but we observe that this choice proves to be more numerically efficient. The counting field modifies the coherent hopping terms according to
\be\label{eq:Hxxztilted}
\hat H_\text{xxz}^{(u)} :=\sum_{j=1}^{N-1}\left[2\varepsilon\left(e^{-\widetilde u}\hat\sigma_{j}^{+}\hat\sigma_{j+1}^{-}+e^{\widetilde u} \hat\sigma_{j}^{-}\hat\sigma_{j+1}^{+}\right)+\Delta\hat\sigma_{j}^{z}\hat\sigma_{j+1}^{z}\right]
\ee
where we set
$\widetilde u:=\frac{u}{2(N-1)}$.
The asymptotic growth rate of $\operatorname{tr} \hat\rho_u(t)$ gives precisely the finite-size cumulant generating function
\begin{equation}
\Psi_N(u)
:=\lim_{t\to\infty}\frac{1}{t}\log\operatorname{tr}\hat\rho_u(t)
=
\lim_{t\to\infty}\partial_t
\log\operatorname{tr}\hat\rho_u(t).
\end{equation}
The rescaled cumulant generating function is then
\begin{equation}
\mathcal F(u)=\lim_{N\to\infty}N\Psi_N(u).
\end{equation}
For faster convergence in time, we evaluate $\partial_t\log\operatorname{tr}\hat\rho_u(t)$ directly at late times.
In Fig.~\ref{fig:ISEPnumerical}(c), we show how the numerical estimate of $\mathcal{F}(u)$ approaches the MFT prediction
and how it deviates from that of the SSEP process with diffusion constant $D_0= 2 \varepsilon^2/\eta_{\text{eff}}$. To further corroborate the convergence with system size, we show in Fig.~\ref{fig:MFT_N_scaling}(c) the $1/N$ behavior of the higher-order corrections to $\mathcal{F}(u)$ predicted by MFT, where quantum corrections to large deviations are expected to appear in addition to the classical ones~\cite{albert2026universalclassicalquantumfluctuations}.

%
%
\clearpage
\onecolumngrid
\newpage

\setcounter{equation}{0}  
\setcounter{figure}{0}
\setcounter{page}{1}
\setcounter{section}{0}    
\renewcommand\thesection{\arabic{section}}    
\renewcommand\thesubsection{\arabic{subsection}}    
\renewcommand{\thetable}{S\arabic{table}}
\renewcommand{\theequation}{S\arabic{equation}}
\renewcommand{\thefigure}{S\arabic{figure}}
\setcounter{secnumdepth}{2}  
\makeatletter
\newcommand{\smtableofcontents}{%
  \section*{Contents}%
  \@starttoc{smtoc}%
}
\newcommand{\smsection}[1]{%
  \section{#1}%
  \addcontentsline{smtoc}{section}{\protect\numberline{\thesection}#1}%
}
\newcommand{\smsubsection}[1]{%
  \subsection{#1}%
  \addcontentsline{smtoc}{subsection}{\protect\numberline{\thesubsection}#1}%
}
\makeatother
\begin{center}
\hypertarget{SM}{\Large \textbf{Supplementary Material}}\\ \ \\
{\large \textbf{\titleinfo}}
\ \\ \ 
\end{center}
The Supplemental Material is organized as follows. In Sec.~\ref{smsec:effective-strong-coupling}, we perform the tree renormalization group of the microscopic model, derive the effective strong-coupling dynamics of the dephased XXZ model, and establish the emergence of the interacting Markov process of ISEP. In Sec.~\ref{smsec:mft}, we derive the nonlinear hydrodynamic transport coefficients and density correlations from both the microscopic dynamics and MFT. Section~\ref{sec:SM-current-MFT} presents the MFT calculation of the current statistics. Finally, Sec.~\ref{app:numerics} describes in more detail the tensor-network simulations, the extraction of the effective parameters, and the numerical convergence checks.

\bigskip
{\smtableofcontents}
\hrulefill

\smsection{Effective strong-coupling limit}\label{smsec:effective-strong-coupling}

\smsubsection{Tree renormalization RG}\label{App:Tree_lvl_RG}

We begin by providing some details on the dimensional analysis or tree renormalization group of the dephased XXZ dynamics. For convenience, we will work with fermions
instead of spins. After a Jordan-Wigner transformation, we obtain
the model: 
\begin{equation}
\frac{d}{dt}\hat{\rho}=-i\left[\hat{H},\hat{\rho}\right]+{\cal L}\left(\hat{\rho}\right),
\end{equation}
\begin{align}
\hat{H}= & \sum_{j}\left(2\varepsilon\left(\hat{c}_{j}^{\dagger}\hat{c}_{j+1}+\hat{c}_{j+1}^{\dagger}\hat{c}_{j}\right)+\Delta\left(2\hat{n}_{j}-1\right)\left(2\hat{n}_{j+1}-1\right)\right),
\end{align}
\begin{equation}
{\cal L}\left(\hat{\rho}\right)=4\eta\sum_{j}\left(\hat{n}_{j}\hat\rho\hat{n}_{j}-\frac{1}{2}\left\{ \hat{n}_{j},\hat\rho\right\} \right).
\end{equation}
The boundary terms are obtained from the spins one by sending $\hat{\sigma}^{+},\hat{\sigma}^{-}\to\hat{c}^{\dagger},\hat{c}$
. 

For simplicity, we discard the chemical potential term and the constant
energy shift in the Hamiltonian: 
\begin{equation}
\hat{H}:=\sum_{j}\left(2\varepsilon\left(\hat{c}_{j}^{\dagger}\hat{c}_{j+1}+\hat{c}_{j+1}^{\dagger}\hat{c}_{j}\right)+4\Delta\hat{n}_{j}\hat{n}_{j+1}\right).
\end{equation}
Since we are dealing with an out-of-equilibrium problem, the field
theory can be expressed using a Keldysh path integral \cite{Kamenev2011,Diehl_KeldyshLindblad}. The generating
function $Z:={\rm tr}\left(\hat\rho_{t}\right)$ can be expressed as 
\begin{align}
Z & =\int\left[d\psi \right]e^{iS\left[\boldsymbol{\psi}\right]}\\
S & =S_{0}+S_{\Delta}+S_{\eta},
\end{align}
with $\boldsymbol{\psi}:=\left\{ \psi^{1},\bar{\psi}^{1},\psi^{2},\bar{\psi}^{2}\right\} $.
$\psi^{1}$ and $\psi^{2}$ are the fields in the Larkin-Ovchinnikov \cite{Larkin_vortices_supra}
basis obtained from the forward and backward fields with the rotation
$\psi^{\pm}=\frac{1}{\sqrt{2}}\left(\psi^{1}\pm\psi^{2}\right)$, 
$\bar{\psi}^{\pm}=\frac{1}{\sqrt{2}}\left(\pm\bar{\psi}^{1}+\bar{\psi}^{2}\right).$
$S_{0}$, $S_{\Delta}$, $S_{\eta}$ designate the action of the kinetic
part, the interacting part, and the dephasing part of the dynamics
respectively. Their explicit expressions are 
\begin{align}
S_{0} & =\int dt\sum_{j}\Big(i\left(\bar{\psi}_{j}^{1}\partial_{t}\psi_{j}^{1}+\bar{\psi}_{j}^{2}\partial_{t}\psi_{j}^{2}\right)-2\varepsilon\left(\bar{\psi}_{j}^{1}\psi_{j+1}^{1}+\bar{\psi}_{j}^{2}\psi_{j+1}^{2}+\bar{\psi}_{j+1}^{1}\psi_{j}^{1}+\bar{\psi}_{j+1}^{2}\psi_{j}^{2}\right),\nonumber \\
S_{\Delta} & =2\Delta\int dt\sum_{j}\Bigg(\bar{\psi}_{j}^{1}\bar{\psi}_{j+1}^{1}\psi_{j}^{2}\psi_{j+1}^{1}+\circlearrowright^{\left(2\right)}+\left(1\right)\leftrightarrow\left(2\right)\Bigg),\\
S_{\eta} & =i4\eta\int dt\sum_{j}\bar{\psi}_{j}^{1}\psi_{j}^{1}\bar{\psi}_{j}^{2}\psi_{j}^{2}.\nonumber 
\end{align}
The notation $\circlearrowright^{\left(2\right)}$ means the sum over
all permutations of index $2$ (3 other terms in this case) and
$\left(1\right)\leftrightarrow\left(2\right)$ means the same terms
but with the indices $1$ and $2$ swapped. Let us now introduce the
lattice spacing $a$ and the continuous fields $\boldsymbol{\psi}_{x=aj}:=\boldsymbol{\psi}_{j}$.
Ignoring the global chemical potential term again, we get, to leading
order in lattice spacing $a$:
\begin{align}
S_{0} & =\int dt\int\frac{d^{d}x}{a^{d}}\left(\bar{\psi}_{x}^{1}\bar{\psi}_{x}^{2}\right)\begin{pmatrix}i\partial_{t}-4\varepsilon a^{2}\partial_{x}^{2}+i0^{+} & 0\\
0 & i\partial_{t}-4\varepsilon a^{2}\partial_{x}^{2}-i0^{+}
\end{pmatrix}\begin{pmatrix}\psi_{x}^{1}\\
\psi_{x}^{2}
\end{pmatrix},\nonumber \\
S_{\Delta} & =2\Delta\int dt\int\frac{d^{d}x}{a^{d}}\Bigg(\bar{\psi}_{x}^{1}\left(1+a\partial_{x}\right)\bar{\psi}_{x}^{1}\psi_{x}^{2}\left(1+a\partial_{x}\right)\psi_{x}^{1}+\circlearrowright^{\left(2\right)}+\left(1\right)\leftrightarrow\left(2\right)\Bigg),\\
S_{\eta} & =i4\eta\int dt\int\frac{d^{d}x}{a^{d}}\bar{\psi}_{x}^{1}\psi_{x}^{1}\bar{\psi}_{x}^{2}\psi_{x}^{2}.\nonumber 
\end{align}
One point worth emphasizing is that we are expanding here the kinetic
term around $k=0$, not linearizing around some Fermi points.
This is appropriate here because we expect the dephasing to destroy
the ballistic Lüttinger-liquid description. 

Let us now see how these terms behave after one renormalization step
from momentum cut-off $\Lambda$ to $\Lambda/b$, $b>1$. Let $\boldsymbol{\psi}_{</>}$
designate the slow and fast fields. After one step of renormalization,
one rescales position, time, and the fields as 
\begin{align}
r & =br',\\
t & =b^{z}t',\\
\psi_{<}\left(br',b^{z}t'\right) & =b^{\chi}\psi'\left(r',t'\right),\\
\bar{\psi}_{<}\left(br',b^{z}t'\right) & =b^{\bar{\chi}}\bar{\psi}\left(r',t'\right).
\end{align}
Keeping the Gaussian part $S_{0}$ of the action invariant imposes
\begin{equation}
d+\chi+\bar{\chi}=0,\quad z=2.
\end{equation}
$\Delta$ then rescales as 
\begin{equation}
\Delta'=\Delta b^{d+2\left(\chi+\bar{\chi}\right)+z}=\Delta b^{2-d},
\end{equation}
and similarly for $\eta$ 
\begin{equation}
\eta':=\eta b^{2-d},
\end{equation}
 which are the results reported in the main text. 

Finally, let us briefly comment on the boundary terms. Their action
would be of the form 
\begin{equation}
S_{{\rm bdy}}\sim\alpha\int dt\bar{\psi}\psi
\end{equation}
So after one-step RG the rescaling becomes: 
\begin{equation}
\alpha'=\alpha b^{2-d}
\end{equation}
and the boundary terms have a similar strong-coupling flow, meaning that
the boundary values are pinned to their equilibrium values in the
thermodynamic limit.

\smsubsection{Derivation of the Markov process}
As established in the previous subsection, in one-dimension $d=1$, the anisotropy parameter $\Delta$ and the dephasing strength $\eta$ are both relevant perturbations with scaling dimension one under the RG flow, $\eta(\ell) \sim\Delta(\ell) \sim e^\ell$. Since the analysis was performed at tree level, it determines only their scaling dimensions, while the corresponding nonuniversal scaling coefficients, $\eta_{\rm eff} =  e^{-\ell}\eta(\ell)$ and $\Delta_{\rm eff} =  e^{-\ell}\Delta(\ell)$, would require a full perturbative treatment beyond leading order.\\
In essence, this argument suggests that the thermodynamic limit of the original dephased XXZ model is equivalent to that of an effective model in which the couplings $\Delta$ and $\eta$ are first taken to be large. More precisely, we consider the rescaled bulk dynamics
\begin{equation}\label{eq:XXZresc}
\partial_t\hat\rho=\left[{\cal L}+e^\ell{\cal L}_b\right](\hat\rho),
\end{equation}
where ${\cal L}$ is the coherent generator introduced below Eq.~\eqref{eq:strongcouplingMaster} of the main text, and
\begin{equation}
{\cal L}_b(\hat\rho)=-\sum_{j=1}^{N_{\rm eff}}\left(\frac{\eta_{\rm eff}}{2}[\hat\sigma_j^z,[\hat\sigma_j^z,\hat\rho]]+i\Delta_{\rm eff}[\hat\sigma_j^z\hat\sigma_{j+1}^z,\hat\rho]\right).
\label{eq:SM-Lb}
\end{equation}
Here, $N_{\rm eff}\gg1$ corresponds to the number of sites in the effective system, so that the size of the original microscopic system is $N=e^\ell N_{\rm eff}$. Introducing the coarse-graining factor $e^\ell$ separately from $N_{\rm eff}$ makes the order of limits explicit: one first takes $e^\ell\to\infty$, thereby reaching the strong-coupling regime of the effective dynamics, and only subsequently considers the thermodynamic limit $N_{\rm eff}\to\infty$.\\

In writing Eq.~\eqref{eq:XXZresc}, we omitted the boundary contributions and may therefore assume periodic boundary conditions for simplicity, as the purpose of this section is to characterize the bulk effective theory in the thermodynamic limit. In the boundary-driven setup considered in the manuscript, the reservoir coupling $\gamma$ also scales linearly under the RG flow, $\gamma(\ell)\sim e^\ell$, as shown in the previous subsection. This drives the boundary sites into a strong-coupling regime in which their magnetization densities are effectively pinned. The resulting dynamics may then be viewed as that of a chain with two fewer sites and modified boundary rates, but with unchanged boundary densities; see Ref.~\cite{Costa_emergence_universality_2026_PRL}. Such microscopic differences vanish in the thermodynamic limit; we therefore disregard the boundary terms in what follows and focus exclusively on the bulk dynamics.\\

In the limit $e^\ell\to\infty$, the interaction and dephasing terms dominate, while coherent hopping is parametrically suppressed, freezing the dynamics at leading order. Following Ref.~\cite{10.21468/SciPostPhys.3.5.033}, second-order perturbation theory in $e^{-\ell}$ applied to Eq.~\eqref{eq:XXZresc} yields an effective evolution on the rescaled timescale $s=e^{-\ell}t$. This evolution is restricted to the slow subspace $\operatorname{Ker}{\cal L}_b\ni\bar\rho_s$ (the kernel of ${\cal L}_b$) and is governed by
\begin{equation}\label{eq:SM-strongcouplingMaster}
\partial_s\bar\rho_s=\mathfrak{L}(\bar\rho_s)
=\left[{\cal P}{\cal L}\left({\cal L}_b^\perp\right)^{-1}{\cal L}{\cal P}\right](\bar\rho_s),
\end{equation}
which is the strong-coupling dynamics reported in Eq.~\eqref{eq:strongcouplingMaster} of the main text. Here, ${\cal P}$ denotes the projector onto $\operatorname{Ker}{\cal L}_b$, the superscript $\perp$ refers to its complementary subspace, and ${\cal L}_b^\perp$ denotes the restriction of ${\cal L}_b$ to that subspace, where it is invertible.

The kernel $\operatorname{Ker}{\cal L}_b$ is spanned by projectors onto classical configurations. The evolving state $\bar\rho_s$ is therefore a classical mixture of the form
\begin{equation}\label{eq:SM-rho-classical}
\bar\rho_s=\sum_{\mathcal C}\Pi_s(\mathcal C)\,|\mathcal C\rangle\langle\mathcal C|.
\end{equation}
where each classical configuration is specified by $\mathcal C=\{\nu_j\}_{j=1}^{N}$, with $\nu_j\in\{-1,1\}$,
\begin{equation}
|\mathcal C\rangle\langle\mathcal C|=\prod_j\hat{\mathbb P}_j^{\nu_j},
\qquad
\hat{\mathbb P}_j^\nu:=\frac{\hat{\mathbb I}+\nu\hat\sigma_j^z}{2}.
\end{equation}
The coefficients $\Pi_s(\mathcal C)$ define a probability distribution over the classical configurations and thus satisfy the normalization condition
$\sum_{\mathcal C}\Pi_s(\mathcal C)=1$. Since the dynamics only takes place in the ``classical space'', it is natural to suppose that a Markov process describes the effective dynamics. Namely, from Eqs.~\eqref{eq:SM-strongcouplingMaster} and \eqref{eq:SM-rho-classical}, we write
\begin{equation}\label{eq:evolprob}
\partial_s\Pi_s(\mathcal C)=\sum_{\mathcal C'} \Pi_s(\mathcal C')\, {\rm tr}\!\left[ \mathfrak{L}(|{\cal C}'\rangle\langle{\cal C}'|)\ |\mathcal C\rangle\langle\mathcal C|\right].
\end{equation}
In order to evaluate the matrix elements in Eq.~\eqref{eq:evolprob}, we make use of the following Pauli identities,
\begin{equation}
\begin{aligned}
[\hat\sigma^\varsigma,\hat{\mathbb P}^\nu]&=-\varsigma\nu\,\hat\sigma^\varsigma, & [\hat\sigma^z,\hat\sigma^\varsigma]&=2\varsigma\,\hat\sigma^\varsigma, &\hat\sigma^\varsigma\hat\sigma^{-\varsigma}&=\hat{\mathbb P}^\varsigma,
\end{aligned}
\qquad
\varsigma,\nu\in\{-1,1\},
\end{equation}
where $\hat\sigma^{+1}:=\hat\sigma^+$ and $\hat\sigma^{-1}:=\hat\sigma^-$. Site indices are omitted here, since operators acting on different sites commute. These relations give
\begin{equation}
{\cal L}(|{\cal C}\rangle\langle {\cal C}|)=-2i\varepsilon\sum_j\left(\hat\sigma_j^{\nu_{j+1}}\hat\sigma_{j+1}^{-\nu_{j+1}}-\hat\sigma_j^{\nu_j}\hat\sigma_{j+1}^{-\nu_j}\right)\prod_{k\neq j,j+1}\hat{\mathbb P}_k^{\nu_k}.
\end{equation}
The off-diagonal operators appearing in this expression are eigenoperators of ${\cal L}_b$:
\begin{align}
{\cal L}_b\!\left[\hat\sigma_j^\varsigma\hat\sigma_{j+1}^{-\varsigma}\prod_{k\neq j,j+1}\hat{\mathbb P}_k^{\nu_k}\right]={}&
-4\eta_{\rm eff}\left[1+\frac{i\Delta_{\rm eff}}{2\eta_{\rm eff}}\varsigma(\nu_{j-1}-\nu_{j+2})\right]\hat\sigma_j^\varsigma\hat\sigma_{j+1}^{-\varsigma}\prod_{k\neq j,j+1}\hat{\mathbb P}_k^{\nu_k}.
\end{align}
Moreover,
\begin{align}
{\cal P}{\cal L}\!\left[\hat\sigma_j^\varsigma\hat\sigma_{j+1}^{-\varsigma}\prod_{k\neq j,j+1}\hat{\mathbb P}_k^{\nu_k}\right]=2\varsigma i\varepsilon\left(\hat\sigma_j^z\hat{\mathbb P}_{j+1}^{-\varsigma}-\hat{\mathbb P}_j^{-\varsigma}\hat\sigma_{j+1}^z\right)\prod_{k\neq j,j+1} \hat{\mathbb P}_k^{\nu_k}.
\end{align}
The action of $({\cal L}_b^\perp)^{-1}$ is therefore obtained by dividing each eigenoperator by its corresponding eigenvalue. Combining the above identities yields
\begin{equation}
\mathfrak{L}(|{\cal C}\rangle\langle{\cal C}|)=\frac{2\varepsilon^2}{\eta_{\rm eff}}\sum_j\frac{|\tau_{j,j+1}\cdot\mathcal C\rangle\langle\tau_{j,j+1}\cdot\mathcal C|-|{\cal C}\rangle\langle {\cal C}|}{1+\left(\frac{\Delta_{\rm eff}}{2\eta_{\rm eff}}\right)^2(\nu_{j+2}-\nu_{j-1})^2},
\end{equation}
where $\tau_{j,j+1}$ exchanges the occupations at sites $j$ and $j+1$, i.e.,
\begin{equation}
\tau_{j,j+1}\cdot \mathcal C= \tau_{j,j+1}\cdot \{\ldots,\nu_j,\nu_{j+1},\ldots\}=\{\ldots,\nu_{j+1},\nu_j,\ldots\}.
\end{equation}
Substituting this into Eq.~\eqref{eq:evolprob} gives the master equation
\begin{equation}\label{eq:SM-Markov-rate}
\partial_s\Pi_s(\mathcal C)=\sum_jW_j(\mathcal C)\left[\Pi_s(\tau_{j,j+1}\cdot\mathcal C)-\Pi_s(\mathcal C)\right],
\qquad
W_j(\mathcal C)=\frac{2\varepsilon^2}{\eta_{\rm eff}}\frac{1}{1+\left(\frac{\Delta_{\rm eff}}{2\eta_{\rm eff}}\right)^2(\nu_{j+2}-\nu_{j-1})^2}.
\end{equation}
Equation~\eqref{eq:SM-Markov-rate} describes a continuous-time Markov process consisting of nearest-neighbor exchanges. The rates $W_j(\mathcal C)$ are non-negative, ensuring that $\Pi_s(\mathcal C)$ remains positive, while the gain and loss terms impose probability conservation. The exchange rate between any two sites $j$ and $j+1$ also depends on the occupations of the two next-nearest neighbors, $j-1$ and $j+2$: it is equal to $2\varepsilon^2/\eta_{\rm eff}$ when their magnetizations coincide and is reduced by a factor $1+(\Delta_{\rm eff}/\eta_{\rm eff})^2$ when they are opposite. Up to higher-order corrections in $e^{-\ell}$, the long-time dynamics is therefore governed by this effective classical stochastic model, whose equivalence with the ISEP formulation of the main text is established in the following subsection.

\smsubsection{Equivalence with the ISEP formulation}
\label{sec:SM-ISEP-connection}

We now connect the Markov process derived above with the ISEP dynamics introduced in the main text. We first relate the two notations used for the configurations. In the derivation above, a spin configuration is denoted by ${\cal C}=\{\nu_j\}_{j=1}^N$, with $\nu_j=\pm1$, while in the main text the corresponding particle configuration is denoted by $\mathbf n=(n_1,\ldots,n_N)$, with
\begin{equation}
\label{eq:JWn}
\hat n_j:=\frac{\hat{\mathbb I}+\hat\sigma_j^z}{2},
\qquad
n_j=\frac{1+\nu_j}{2},
\end{equation}
such that $\nu_j=+1$ and $\nu_j=-1$ correspond, respectively, to an occupied and an empty site. The exchange rate in Eq.~\eqref{eq:SM-Markov-rate} depends only on the spin configurations at the two outer sites $j-1$, $j+2$. It can therefore be parametrized by the four amplitudes (cf. Eq.~\eqref{eq:A-ampl} of the main text)
\begin{equation}
A_{\uparrow,\uparrow} = A_{\downarrow,\downarrow} =\frac{2\varepsilon^2}{\eta_\text{eff}}, \qquad A_{\uparrow,\downarrow} = A_{\downarrow,\uparrow}=\frac{2\varepsilon^2/\eta_{\rm eff}}{1+\left(\frac{\Delta_\text{eff}}{\eta_\text{eff}}\right)^2}.
\label{eq:SM-A-ampl}
\end{equation}
The first rate applies when the two outer spins coincide, whereas the second applies when they are opposite. Introducing also
\begin{equation}
\hat{\mathbf P}_j^{\alpha,\beta}:=\sqrt{A_{\alpha,\beta}}\, \hat{\mathbb P}_{j-1}^{\alpha} \hat{\mathbb P}_{j+2}^{\beta};
\qquad
\hat{\mathbb P}^{\uparrow}_j=\hat n_j;
\qquad
\hat{\mathbb P}^{\downarrow}_j=1-\hat n_j,
\end{equation}
the jump operators of the main text take the form
\begin{equation}
\hat L_{j;\pm}^{\alpha,\beta}= \hat{\mathbf P}_j^{\alpha,\beta} \hat\sigma_j^\pm\hat\sigma_{j+1}^\mp.
\label{eq:SM-conditioned-jumps}
\end{equation}
Thus, the four possible values of $\alpha,\beta\in\{\uparrow,\downarrow\}$ specify the occupations of the outer sites, while the index $\pm$ specifies the two hopping directions across the link $(j,j+1)$. The effective dynamics is therefore
\begin{equation}
\partial_s\hat\rho=\sum_j \sum_{\alpha,\beta\in\{\uparrow,\downarrow\}} \left[{\cal D}_{\hat L_{j;+}^{\alpha,\beta}}(\hat\rho)+{\cal D}_{\hat L_{j;-}^{\alpha,\beta}}(\hat\rho)\right]+{\cal L}_\text{bdy}(\hat\rho),
\label{eq:SM-effective-Lindblad}
\end{equation}
 which is Eq.~\eqref{eq:eff-dyn} of the main text. 
Eq.~\eqref{eq:SM-effective-Lindblad} reduces to a classical nearest-neighbor exchange process. The rate associated with the link $(j,j+1)$ is the diagonal matrix element of
\begin{equation}
\hat D_j:= \sum_{\alpha,\beta} \left(\hat{\mathbf P}_j^{\alpha,\beta}\right)^2= \sum_{\alpha,\beta} A_{\alpha,\beta} \hat{\mathbb P}_{j-1}^{\alpha} \hat{\mathbb P}_{j+2}^{\beta}.
\label{eq:SM-Dj-projectors}
\end{equation}
Using Eq.~\eqref{eq:SM-A-ampl}, this becomes
\begin{align}
\hat D_j=& A_{\downarrow,\downarrow}+\left(A_{\uparrow,\downarrow}-A_{\downarrow,\downarrow}\right)\hat n_{j-1}+\left(A_{\downarrow,\uparrow}-A_{\downarrow,\downarrow}\right)\hat n_{j+2}+ \left(A_{\uparrow,\uparrow}+A_{\downarrow,\downarrow}-A_{\uparrow,\downarrow}-A_{\downarrow,\uparrow}\right)\hat n_{j-1}\hat n_{j+2}
\nonumber\\
=& D_0\left[ 1+\lambda\hat n_{j-1}(1-\hat n_{j+2})+\lambda\hat n_{j+2}(1-\hat n_{j-1})\right],
\label{eq:SM-Dj-ISEP}
\end{align}
where
\begin{equation}
D_0:=A_{\downarrow,\downarrow}=\frac{2\varepsilon^2}{\eta_{\rm eff}},
\qquad
\lambda:=\frac{A_{\uparrow,\downarrow}-A_{\downarrow,\downarrow}}{A_{\downarrow,\downarrow}}=-\left[1+\left(\frac{\eta_{\rm eff}}{\Delta_{\rm eff}}\right)^2\right]^{-1}.
\label{eq:SM-D0-lambda}
\end{equation}
The transition rates associated with the Markov process derived above are therefore
\begin{equation}
W_j({\cal C})=\langle\mathbf n|\hat D_j|\mathbf n\rangle.
\end{equation}
Resolving the exchange into the two possible hopping directions gives the transition rates 
\begin{equation}
\underset{j\to j+1}{f(\mathbf n)}= \langle\mathbf n|\hat D_j|\mathbf n\rangle n_j(1-n_{j+1}),
\qquad
\underset{j+1\to j}{f(\mathbf n)}=\langle\mathbf n|\hat D_j|\mathbf n\rangle n_{j+1}(1-n_j).
\end{equation}
The master equation \eqref{eq:SM-Markov-rate} can then be expressed as
\begin{align}
\partial_s \Pi_s(\mathbf{n})=\sum_{j=1}^{N-2} \Big(&\underset{j\to j+1}{f(\mathbf{n})}\Pi_s(\mathbf{n}^{j\to j+1})+\underset{j+1\to j}{f(\mathbf{n})}\Pi_s(\mathbf{n}^{j+1\to j}) -[\underset{j\to j+1}{f(\mathbf{n})}+\underset{j+1\to j}{f(\mathbf{n})}]\Pi_s(\mathbf{n})\Big)
\end{align}
where the configurations $\mathbf{n}^{j\to k}$ are obtained by moving a particle from site $j\to k$. This establishes the equivalence between the Markov process derived in the previous subsection and the ISEP formulation of the main text.\\

For later use, we finally express the dynamics of density observables in the Heisenberg picture. We introduce
\begin{equation}
\hat\ell_j:= \hat\sigma_j^-\hat\sigma_{j+1}^+=\hat c_{j+1}^\dagger\hat c_j,
\end{equation}
with $\hat c^\dagger_j$ and $\hat c_j$ standard spinless-fermion operators, such that
\begin{equation}
\hat L_{j;-}^{\alpha,\beta}=\hat{\mathbf P}_j^{\alpha,\beta}\hat\ell_j,
\qquad
\hat L_{j;+}^{\alpha,\beta}=\hat{\mathbf P}_j^{\alpha,\beta}\hat\ell_j^\dagger.
\end{equation}
For any operator $\hat O_{\hat n}=\hat n_{i_1}\dots \hat n_{i_k}$ made only of local densities at distinct sites, one has
\begin{equation}
[\hat{\mathbf P}_j^{\alpha,\beta},\hat O_{\hat n}]=0; \quad [\hat{\mathbf P}_j^{\alpha,\beta},\hat\ell_j]=[\hat{\mathbf P}_j^{\alpha,\beta},\hat\ell_j^\dagger]=0,
\end{equation}
since $\hat{\mathbf P}_j^{\alpha,\beta}$ is supported on sites $j-1$, $j+2$. One then finds
\begin{align}
&{\cal D}_{\hat L_{j;-}^{\alpha,\beta}}^*(\hat O_{\hat n})+{\cal D}_{\hat L_{j;+}^{\alpha,\beta}}^*(\hat O_{\hat n})= \left(\hat{\mathbf P}_j^{\alpha,\beta}\right)^2\left[{\cal D}_{\hat \ell_j}(\hat O_{\hat n})+{\cal D}_{\hat \ell_j^\dagger}(\hat O_{\hat n})\right].
\end{align}
Summing over $(\alpha,\beta)$ and taking the expectation $\langle\bullet\rangle:={\rm tr}(\bar\rho\ \bullet)$, one obtains the equation of motion (cf. Eq.~\eqref{eq:EM-Heisenberg} of the \hyperlink{EM}{End Matter})
\be\label{eq:eom}
\partial_s\langle\hat O_{\hat n}\rangle=\sum_{j} \langle \hat D_j\left[{\cal D}_{\hat \ell_j}(\hat O_{\hat n})+  {\cal D}_{\hat \ell^\dagger_j}(\hat O_{\hat n})\right]\rangle +\langle {\cal L}_\text{bdy}^*(\hat O_{\hat n})\rangle.
\ee
This is the starting point for the hydrodynamic equations derived below.

\smsection{Details on the ISEP and MFT computations}
\label{smsec:mft}
\smsubsection{Microscopic derivation of the nonlinear hydrodynamics and diffusivity}
\label{sec:SM-density}
We first provide further details on the emergence of the nonlinear diffusion equation from the effective dynamics in Eq.~\eqref{eq:eom}. For density operators, it is convenient to introduce the symmetric exclusion generator
\begin{equation}
{\cal L}^\text{ssep}_j(\bullet):={\cal D}_{\hat \ell_j}(\bullet)+{\cal D}_{\hat \ell_j^\dagger}(\bullet),
\end{equation}
such that the bulk equation of motion reads
\begin{equation}
\partial_s\langle\hat O_{\hat n}\rangle=\sum_j\left\langle\hat D_j{\cal L}^\text{ssep}_j(\hat O_{\hat n})\right\rangle+\langle{\cal L}_\text{bdy}^*(\hat O_{\hat n})\rangle.
\label{eq:SM-density-generator}
\end{equation}
Applying Eq.~\eqref{eq:SM-density-generator} to the local density $\hat n_j$ gives the exact lattice conservation law
\begin{equation}
\partial_s\langle\hat n_j\rangle=\langle\hat J_{j-1}-\hat J_j\rangle+\langle{\cal L}_\text{ bdy}^*(\hat n_j)\rangle,
\label{eq:SM-lattice-continuity}
\end{equation}
with microscopic current
\begin{equation}
\hat J_j:=-\hat D_j\left(\hat n_{j+1}-\hat n_j\right).
\label{eq:SM-microscopic-current}
\end{equation}
Away from the boundaries, the current expectation value can be rewritten as
\begin{align}
\langle\hat J_j\rangle
=&-\langle\hat D_j\rangle\left(\langle\hat n_{j+1}\rangle-\langle\hat n_j\rangle\right)
-\left\langle\hat D_j\left(\hat n_{j+1}-\hat n_j\right)\right\rangle^c
\nonumber\\
=&-\langle\hat D_j\rangle\left(\langle\hat n_{j+1}\rangle-\langle\hat n_j\rangle\right)
+{\cal O}(N^{-2}).
\label{eq:SM-current-decomposition}
\end{align}
The MFT scaling in Eq.~\eqref{eq:MFT_scaling_density} of the main text implies that $\langle\hat D_j\hat n_j\rangle^c={\cal O}(N^{-1})$ since $\hat D_j$ is supported on sites $j+2,j-1$. The additional discrete gradient in the second term of Eq.~\eqref{eq:SM-current-decomposition} therefore makes this contribution subleading. One thus finds, to leading order,
\be
\partial_s\langle\hat n_j\rangle\simeq\nabla_j^-\left(\langle\hat D_j\rangle\nabla_j^+\langle\hat n_j\rangle\right)+\langle{\cal L}_\text{bdy}^*(\hat n_j)\rangle,
\ee
with discrete derivatives $\nabla_j^+f_j:=f_{j+1}-f_j$ and $\nabla_j^-f_j:=f_j-f_{j-1}$. For the interaction in Eq.~\eqref{eq:interaction} of the main text,
\begin{equation}
\hat D_j=D_0\left[1+\lambda\hat n_{j-1}(1-\hat n_{j+2})+\lambda\hat n_{j+2}(1-\hat n_{j-1})\right],
\end{equation}
the MFT factorization gives
\begin{align}
\langle\hat D_j\rangle
=&D_0\Big[1+\lambda\bar n_{j-1}(1-\bar n_{j+2})+\lambda\bar n_{j+2}(1-\bar n_{j-1})\Big]
+{\cal O}(N^{-1})
\nonumber\\
=&D(\bar n_x)+{\cal O}(N^{-1}),
\end{align}
where
\be
\bar n_x:=\lim_{N\to\infty}\langle\hat n_j\rangle\Big\vert_{x=j/N}
\ee
and
\begin{equation}
D(\bar n_x):=\lim_{N\to\infty}\langle\hat D_j\rangle\Big\vert_{x=j/N}
=D_0\left[1+2\lambda\bar n_x(1-\bar n_x)\right].
\label{eq:SM-diffusivity}
\end{equation}

Introducing the diffusive time $\tau=s/N^2$ and the rescaled average current
\begin{equation}
\bar J_x:=\lim_{N\to\infty}N\langle\hat J_j\rangle\Big\vert_{x=j/N},
\end{equation}
Eqs.~\eqref{eq:SM-lattice-continuity} and \eqref{eq:SM-current-decomposition} yield
\begin{align}
\label{eq:effec-Current-contlimit}
\partial_\tau\bar n_x&=-\partial_x\bar J_x,
\nonumber\\
\bar J_x&=-D(\bar n_x)\partial_x\bar n_x,
\end{align}
or, equivalently,
\begin{equation}
\partial_\tau\bar n_x=\partial_x\left[D(\bar n_x)\partial_x\bar n_x\right],
\label{eq:SM-nonlinear-diffusion}
\end{equation}
with boundary conditions
\begin{equation}
\bar n_{0}=\frac{1-\mu}{2}=:n_a;
\qquad
\bar n_{1}=\frac{1+\mu}{2}=:n_b.
\label{eq:SM-density-boundaries}
\end{equation}

In the steady state, the current is constant, $\bar J_x:=\bar J$, and Eq.~\eqref{eq:SM-nonlinear-diffusion} can be integrated once. Introducing the primitive
\begin{equation}
\Phi(n):=\int^n dq\,D(q),
\end{equation}
the stationary profile is determined implicitly by
\begin{equation}
\Phi(\bar n_x)=(1-x)\Phi(n_a)+x\,\Phi(n_b),
\label{eq:SM-stationary-profile}
\end{equation}
while the stationary current is
\begin{equation}
\bar J=-\left[\Phi(n_b)-\Phi(n_a)\right].
\label{eq:SM-stationary-current}
\end{equation}
For the diffusivity in Eq.~\eqref{eq:SM-diffusivity}, a convenient choice of primitive is
\begin{equation}\label{eq:SM-Phi}
\Phi(n)=D_0\left[n+\lambda n^2-\frac{2\lambda}{3}n^3\right].
\end{equation}
The stationary density profile is then obtained by inverting the cubic polynomial in Eq.~\eqref{eq:SM-stationary-profile}. In particular, we find the following analytical expression
\begin{equation}
 \bar n_x
 =\frac12+\frac{1}{r}
 \sinh\!\left[
   \frac13\operatorname{arsinh}\!\left(
     r\left[3 Y_1(x)+4r^2 Y_3(x)\right]
   \right)
 \right].
 \label{eq:theoretical-density-profile}
\end{equation}
with
\begin{equation}
 r:=\frac{1}{\sqrt{1+2(\eta_{\rm eff}/\Delta_{\rm eff})^2}}
 \label{eq:profile-parameters}
\end{equation}
and
\begin{equation}
  Y_p(x)
 :=(1-x)\left(n_a-\frac12\right)^p
 +x\left(n_b-\frac12\right)^p,
 \qquad p\in\{1,3\}.
 \label{eq:Yp-definition}
\end{equation}
It satisfies $\bar n_0=n_a$ and $\bar n_1=n_b$ and in the non-interacting limit
$\Delta_{\rm eff}\to0$ or equivalently $r\to 0$, it reduces to the
SSEP profile
\begin{equation}
 \bar n_x=(1-x)n_a+xn_b.
\end{equation}
\smsubsection{MFT description of density-density correlations}
\label{sec:SM-MFT-correlations}
We next derive the equations governing density correlations directly within MFT. The starting point is the fluctuating hydrodynamics specified by
\begin{align}
\partial_\tau\n_x&=-\partial_x\jj_x;
\nonumber\\
\jj_x&=-D(\n_x)\partial_x\n_x
+\sqrt{\frac{\sigma(\n_x)}{N}}\,\xi_x,
\label{eq:SM-fluctuating-hydrodynamics}
\end{align}
where, we recall, $\xi_x$ is a Gaussian space-time white noise satisfying
\begin{equation}\label{eq:white-noise}
\mathbb{E}[\xi_x(\tau)]=0; \quad \mathbb{E}[\xi_x(\tau)\xi_y(\tau')]=\delta(x-y)\delta(\tau-\tau'),
\end{equation}
and $\mathbb{E}[\bullet]$ denotes the statistical average over the Gaussian fluctuations. The density field obeys fixed boundary conditions,
\begin{equation}
\n_{x=0}=n_a,\quad \n_{x=1}=n_b.
\end{equation}
To leading order in $1/N$, averaging Eq.~\eqref{eq:SM-fluctuating-hydrodynamics} reproduces the nonlinear diffusion equation~\eqref{eq:SM-nonlinear-diffusion}. \\

To describe the leading fluctuations, we write
\begin{equation}
\n_x(\tau)=\bar n_x(\tau)+\frac{\delta\n_x(\tau)}{\sqrt{N}}.
\label{eq:SM-density-expansion}
\end{equation}
Linearizing Eq.~\eqref{eq:SM-fluctuating-hydrodynamics} around $\bar n_x$ gives
\begin{equation}
\partial_\tau\delta\n_x=\partial_x^2\left[D(\bar n_x)\delta\n_x\right]-\partial_x\left[\sqrt{\sigma(\bar n_x)}\,\xi_x\right].
\label{eq:SM-linearized-density}
\end{equation}
The form of the deterministic term follows from
\begin{equation}
D(\bar n_x)\partial_x\delta\n_x+D'(\bar n_x)\delta\n_x\partial_x\bar n_x
=\partial_x\left[D(\bar n_x)\delta\n_x\right].
\end{equation}
We then introduce the connected density-density correlation
\begin{equation}
C_{x,y}(\tau):=\mathbb{E}\left[\delta\n_x(\tau)\delta\n_y(\tau)\right].
\label{eq:SM-C-definition}
\end{equation}
Using Eq.~\eqref{eq:SM-linearized-density} and the properties of the white noise~\eqref{eq:white-noise}, one obtains Eq.~\eqref{eq:eq-for-C} of the main text, namely
\begin{equation}
\left(\de_\tau-\Delta_x^D-\Delta_y^D\right)C_{x,y}
=\de_x\de_y\left[\sigma(\bar n_x)\delta(x-y)\right],
\label{eq:SM-MFT-C}
\end{equation}
with generalized Laplacian $\Delta_x^D\bullet:=\partial_x^2\left[D(\bar n_x)\bullet\right]$. Note that $\Delta_x^D$ differs from the diffusive operator $\de_x\left[D(\bar n_x)\de_x\bullet\right]$ appearing in Eq.~\eqref{eq:SM-nonlinear-diffusion} for the average density. The correlation satisfies Dirichlet boundary conditions,
\begin{equation}
C_{0,y}=C_{1,y}=C_{x,0}=C_{x,1}=0.
\label{eq:SM-C-boundaries}
\end{equation}

Importantly, the correlation in Eq.~\eqref{eq:SM-MFT-C} is understood as a distribution. It can be decomposed into a contact contribution at $x=y$ and a regular non-equilibrium part,
\begin{equation}
C_{x,y}=\chi(\bar n_x)\delta(x-y)+C^\text{neq}_{x,y},
\label{eq:SM-C-decomposition}
\end{equation}
where $\chi(n)$ is the static compressibility. At equilibrium, where $\bar n_x=\bar n$ is constant, Eq.~\eqref{eq:SM-MFT-C} gives
\begin{equation}
\chi(\bar n)=\frac{\sigma(\bar n)}{2D(\bar n)}.
\label{eq:SM-Einstein}
\end{equation}
For an exclusion process, local equilibrium is described by the Bernoulli measure, and therefore
\begin{equation}
\chi(n)=n(1-n).
\end{equation}
The fluctuation-dissipation relation then gives
\begin{equation}
\sigma(n)=2D(n)n(1-n).
\label{eq:SM-MFT-mobility}
\end{equation}
Below, we recover this relation directly from the microscopic ISEP dynamics.

\subsection{Microscopic derivation of the density-density cumulant and mobility}
\label{sec:SM-microscopic-correlations}

We now derive Eq.~\eqref{eq:SM-MFT-C} for the scaling function of the density-density cumulant $\langle \hat n_i\hat n_j\rangle^c$ from the microscopic equation of motion~\eqref{eq:eom}. For the sake of clarity, we introduce the following notation for cumulants
\be
\langle \hat A;\hat B\rangle^c:=\langle\hat A\hat B\rangle-\langle \hat A\rangle\langle\hat B\rangle:=\langle \hat A\hat B\rangle^c,
\ee
with $\hat A$, $\hat B$ some generic operators, and with straightforward generalization to higher-order cumulants. In this notation, the grouping of the density operators entering a cumulant remains unambiguous when deriving its dynamics. For bulk sites, a direct application of Eq.~\eqref{eq:SM-density-generator} gives
\begin{align}
\partial_s \langle \hat n_i ; \hat n_j\rangle^c =&\left\langle(\hat J_{i-1}-\hat J_i);\hat n_j\right\rangle^c +\left\langle(\hat J_{j-1}-\hat J_j);\hat n_i\right\rangle^c
\nonumber\\
&-(\delta_{i-1,j}-\delta_{i,j})\left\langle\hat D_{i-1}(\hat n_i-\hat n_{i-1})^2\right\rangle-(\delta_{i,j-1}-\delta_{i,j})\left\langle\hat D_i(\hat n_{i+1}-\hat n_i)^2\right\rangle.
\label{eq:SM-microscopic-C}
\end{align}
The first line describes the diffusive propagation, while the last contains the discrete contact terms. \\

We first consider macroscopically separated points, namely $|i-j|\gg 1$. Using the expression~\eqref{eq:SM-microscopic-current}, one has
\be
\langle-\hat J_i; \hat n_j\rangle^c =\langle\hat D_i\rangle\left\langle(\hat n_{i+1}-\hat n_i);\hat n_j\right\rangle^c +\big(\langle\hat n_{i+1}\rangle-\langle\hat n_i\rangle\big) \langle\hat D_i;\hat n_j\rangle^c +\left\langle\hat D_i;\hat n_{i+1}-\hat n_i;\hat n_j\right\rangle^c.
\label{eq:SM-current-cumulant}
\ee
The last term is of order $N^{-3}$ due to the MFT scaling~\eqref{eq:MFT_scaling_density}. Indeed, the connected three-point function contributes a factor $N^{-2}$, while the discrete gradient gives one additional power of $N^{-1}$.\\

For the second term in Eq.~\eqref{eq:SM-current-cumulant}, the explicit expression of $\hat D_i$ gives
\begin{align}
\langle\hat D_i;\hat n_j\rangle^c =&D_0\lambda\Big[ (1-2\bar n_{i+2})\langle\hat n_{i-1};\hat n_j\rangle^c +(1-2\bar n_{i-1})\langle\hat n_{i+2};\hat n_j\rangle^c\Big]+{\cal O}(N^{-2})
\nonumber\\
=&\frac{1}{N}D'(\bar n_x)C^\text{neq}_{x,y}+{\cal O}(N^{-2}),
\label{eq:SM-Dn-correlation}
\end{align}
where we defined, for $x\neq y$, the regular scaling function
\begin{equation}
C^\text{neq}_{x,y}:=\lim_{N\to\infty}N\langle\hat n_i;\hat n_j\rangle^c
\Big\vert_{\substack{x=i/N\\y=j/N}},
\end{equation}
and $D'(n)=2D_0\lambda(1-2n)$. Similarly,
\begin{equation}
\left\langle(\hat n_{i+1}-\hat n_i);\hat n_j\right\rangle^c =\frac{1}{N^2}\partial_x C^\text{neq}_{x,y}+{\cal O}(N^{-3}).
\end{equation}
Equation~\eqref{eq:SM-current-cumulant} therefore becomes
\begin{equation}
\langle-\hat J_i;\hat n_j\rangle^c =\frac{1}{N^2}\partial_x\left[D(\bar n_x)C^\text{neq}_{x,y}\right]+{\cal O}(N^{-3}).
\label{eq:SM-current-C-continuum}
\end{equation}
Taking the remaining discrete gradient in Eq.~\eqref{eq:SM-microscopic-C} yields, in the diffusive scaling regime $\tau=s/N^2$,
\begin{equation}
\partial_\tau C^\text{neq}_{x,y} =\Delta_x^D C^\text{neq}_{x,y}+\Delta_y^D C^\text{neq}_{x,y},
\qquad x\neq y.
\label{eq:SM-C-away-contact}
\end{equation}

A direct evaluation of the first two terms in Eq.~\eqref{eq:SM-microscopic-C} for microscopic separations $|i-j|={\cal O}(1)$ shows that their continuum limit is still represented by the action of $\Delta_x^D+\Delta_y^D$ on the full correlation $C_{x,y}$, without generating additional contact sources. One therefore has
\be\label{eq:SM-C-diff}
\left\langle(\hat J_{i-1}-\hat J_i);\hat n_j\right\rangle^c +\left\langle(\hat J_{j-1}-\hat J_j);\hat n_i\right\rangle^c \simeq \frac{1}{N^3}\left(\Delta_x^D+\Delta_y^D\right)C_{x,y}.
\ee

It remains to determine the contact source originating from the last two terms in Eq.~\eqref{eq:SM-microscopic-C}. \\

Its physical origin can be understood from the local fluctuations of the integrated current across a bond. During a time step $ds$, a particle can hop across the bond $(i,i+1)$ either from $i$ to $i+1$ or in the opposite direction, with rates
\begin{equation}
\langle\mathbf{n}|\hat D_i\,\hat\ell_i^\dagger\hat\ell_i|\mathbf{n}\rangle =\langle\mathbf{n}|\hat D_i\hat n_i(1-\hat n_{i+1})|\mathbf{n}\rangle,
\qquad
\langle\mathbf{n}|\hat D_i\,\hat\ell_i\hat\ell_i^\dagger|\mathbf{n}\rangle =\langle\mathbf{n}|\hat D_i\hat n_{i+1}(1-\hat n_i)|\mathbf{n}\rangle,
\end{equation}
respectively. These two processes contribute with opposite signs to the integrated current but add in its variance. Defining the local current activity as
\begin{align}
\hat q_i &:=\hat D_i\left[ \hat n_i(1-\hat n_{i+1}) +\hat n_{i+1}(1-\hat n_i)\right]=\hat D_i(\hat n_{i+1}-\hat n_i)^2,
\label{eq:SM-current-activity}
\end{align}
the contribution of local current fluctuations to the density-density cumulant is localized at coinciding sites and enters the microscopic equation as a discrete double gradient of $\langle\hat q_i\rangle$. Since $\hat D_i$ is supported on sites $i-1$ and $i+2$, the MFT factorization gives
\begin{align}
\langle\hat q_i\rangle 
=2D(\bar n_x)\bar n_x(1-\bar n_x)+{\cal O}(N^{-1}).
\label{eq:SM-current-activity-continuum}
\end{align}
This identifies the mobility as
\begin{equation}
\sigma(\bar n_x)=2D(\bar n_x)\bar n_x(1-\bar n_x),
\end{equation}
in agreement with the MFT result~\eqref{eq:SM-MFT-mobility}.\\

Equivalently, the last two terms in Eq.~\eqref{eq:SM-microscopic-C} can be evaluated directly. Using $\delta_{i,j}\to N^{-1}\delta(x-y)$ as $N\to\infty$, one finds, to leading order in $1/N$,
\begin{align}\label{eq:SM-micro-contact}
&-(\delta_{i-1,j}-\delta_{i,j}) \left\langle\hat D_{i-1}(\hat n_i-\hat n_{i-1})^2\right\rangle -(\delta_{i,j-1}-\delta_{i,j}) \left\langle\hat D_i(\hat n_{i+1}-\hat n_i)^2\right\rangle\simeq
\frac{1}{N^3}\partial_x\partial_y\left[\sigma(\bar n_x)\delta(x-y)\right].
\end{align}
Combining Eqs.~\eqref{eq:SM-C-diff} and \eqref{eq:SM-micro-contact}, and using
$\partial_s\langle\hat n_i;\hat n_j\rangle^c=N^{-3}\partial_\tau C_{x,y}$,
one recovers the MFT equation~\eqref{eq:SM-MFT-C} from the microscopic dynamics. This also shows that the interaction modifies the diffusivity and mobility by the same multiplicative factor, while leaving the equilibrium compressibility unchanged.

\smsection{MFT calculation of current statistics}
\label{sec:SM-current-MFT}

We finally derive the MFT prediction for the full-counting statistics of the integrated current. Let $Q_{s,j}$ denote the net charge transferred from left to right across the link $j$ during the time interval $[0,s]$. At finite $N$, its statistics can be defined directly from the effective Markov process by introducing the charge-resolved probability $\Pi_s(\mathbf n,Q)$ that the system is in configuration $\mathbf n$ at time $s$, having transferred a net charge $Q$. Its marginals satisfy
\begin{equation}
\Pi_s(\mathbf n)=\sum_{Q\in\mathbb Z}\Pi_s(\mathbf n,Q), \quad P_s(Q)=\sum_{\mathbf n}\Pi_s(\mathbf n,Q).
\label{eq:SM-charge-resolved-probability}
\end{equation}
The corresponding tilted probability is
\be
\widetilde\Pi_s(u;\mathbf n):= \sum_{Q\in\mathbb Z}e^{uQ}\Pi_s(\mathbf n,Q),
\label{eq:SM-tilted-probability}
\ee
and evolves under a tilted Markov generator in which transitions increasing or decreasing $Q$ by one unit are weighted by $e^u$ or $e^{-u}$, respectively. The same construction can be implemented directly at the level of the effective Lindblad dynamics~\eqref{eq:eff-dyn}. Introducing the tilted density matrix
\be
\hat\rho_u(s):=\sum_{\mathbf n,Q} e^{uQ}\Pi_s(\mathbf n,Q) |\mathbf n\rangle\langle\mathbf n|,
\label{eq:SM-tilted-density-matrix}
\ee
the associated tilted generator ${\cal L}_u$ weights the jumps with the counting field, transferring one particle in the positive and negative directions across the counted link $j$. Explicitly, its contribution at the counted link reads
\be
{\cal L}_{u}^{(j)}(\hat\rho_u)= \hat D_{j} \Big(e^u\hat\ell_{j} \hat\rho_u \hat\ell_{j}^\dagger + e^{-u}\hat \ell_{j}^\dagger  \hat\rho_u \hat \ell_{j}-\frac{1}{2} \left\{\hat\ell_{j}\hat\ell_{j}^\dagger + \hat \ell^\dagger_{j}\hat\ell_{j},\hat\rho_u\right\}\Big),
\label{eq:SM-tilted-Lindblad}
\ee
while all uncounted terms remain unchanged. At $u=0$, the usual trace-preserving dynamics is recovered. For $u\neq0$, the tilted evolution is no longer trace-preserving, and its trace gives the moment generating function
\begin{equation}
Z_s(u):={\rm tr}\big(\hat\rho_u(s)\big) = \sum_{Q\in\mathbb Z}P_s(Q)e^{uQ}=\sum_{\mathbf n}\widetilde\Pi_s(u;\mathbf n).
\label{eq:SM-microscopic-MGF}
\end{equation}
The finite-size cumulant generating function is consequently given by the dominant eigenvalue of the tilted generator,
\begin{equation}\Psi_N(u):=\lim_{s\to\infty}\frac{1}{s}\log Z_s(u).
\label{eq:SM-finite-CGF}
\end{equation}
One can then study the tilted equation of motion and derive the associated cumulant generating function, as done, e.g., in Refs.~\cite{Costa_emergence_universality_2026_PRL,albert2026universalclassicalquantumfluctuations}.\\

Here, we instead proceed differently and directly derive the MFT expression for this quantity. This derivation is standard and can be found, e.g., in Refs.~\cite{Bodineau2010,Derrida2011,Derrida_2025}. On diffusive scales $\tau=s/N^2$, the integrated current through a position $x$ reads
\begin{equation}
{\cal Q}_{\tau,x}=N\int_0^\tau  d\tau'\,\jj_x(\tau')+{\cal O}(N).
\label{eq:SM-integrated-current}
\end{equation}
where ${\cal Q}_{\tau,x}:= Q_{s=\tau N^2,j=xN}$. In the stationary regime, the choice of $x$ does not affect the long-time cumulant generating function. Indeed, by the continuity equation, integrated currents through different sections differ only by the variation of the total charge contained between them, which remains bounded in time. Equivalently, one may use the spatially averaged current,
\begin{equation}
{\cal Q}_{\tau,x}\overset{\tau\gg1}{\approx} {\cal Q}_{\tau}=N\int_0^\tau d\tau'\int_0^1 dx\,\jj_x(\tau')+{\cal O}(N).
\label{eq:SM-space-averaged-current}
\end{equation}

Starting from the fluctuating hydrodynamics~\eqref{eq:SM-fluctuating-hydrodynamics}, the MFT framework assigns to a history $(\n,\jj)$ the probability
\begin{equation}
{\rm Prob}[\n,\jj] \asymp \exp\left[-N{\cal S}[\n,\jj]\right] \delta\left(\partial_{\tau'}\n+\partial_x\jj\right),
\end{equation}
with MFT action
\begin{equation}
{\cal S}[\n,\jj]=\int_0^\tau d\tau'\int_0^1 dx\, \frac{\left[\jj+D(\n)\partial_x\n\right]^2}{2\sigma(\n)}.
\label{eq:SM-MFT-action}
\end{equation}
Here, $\asymp$ denotes logarithmic asymptotic equivalence. Using Eq.~\eqref{eq:SM-space-averaged-current}, the generating function~\eqref{eq:SM-microscopic-MGF} can be expressed as the path integral
\be
Z_{\tau}(u)=\mathbb{E}\left[e^{u{\cal Q}_\tau}\right] \asymp \int [d\n]\,[d\jj]\, \delta\left(\partial_{\tau'}\n+\partial_x\jj\right) \exp\left\{ -N{\cal S}[\n,\jj] +Nu\int_0^\tau d\tau'\int_0^1 dx\,\jj\right\}.
\label{eq:SM-current-path-integral}
\ee
Following standard practice, we then enforce the continuity equation through a response field $p$ and integrate over the Gaussian current field. This gives
\begin{equation}
Z_{\tau}(u) \asymp \int [d\n]\,[dp]\, \exp\left[ -N\int_0^\tau d\tau'\int_0^1 dx \left( p\de_{\tau'}\n-{\cal H}_u[\n,p]\right)\right],
\label{eq:SM-Hamilton-path-integral}
\end{equation}
with biased Hamiltonian density
\be
{\cal H}_u[\n,p]=-D(\n)\partial_x\n\left(\partial_xp+u\right)+\frac{\sigma(\n)}{2}\left(\partial_xp+u\right)^2.
\label{eq:SM-biased-Hamiltonian}
\end{equation}
In this convention, the response field has boundary conditions $p(0,\tau')=p(1,\tau')=0$, while the density remains fixed by the reservoirs, $\n(0,\tau')=n_a$, $\n(1,\tau')=n_b$. Equivalently, the transformation $p\to \tilde p=p+ux$ removes the counting field from the Hamiltonian and yields the boundary conditions $\tilde p(0,\tau')=0$, $\tilde p(1,\tau')=u$.\\

The Hamilton equations, also known as the MFT equations, following from Eq.~\eqref{eq:SM-Hamilton-path-integral} are
\begin{align}
\partial_\tau\n&= \partial_x\left[ D(\n)\partial_x\n -\sigma(\n)(\partial_xp+u)\right];
\label{eq:SM-MFT-n-equation}
\\
\partial_\tau p&=-D(\n)\partial_x^2p-\frac{\sigma'(\n)}{2}\left(\partial_xp+u\right)^2.
\label{eq:SM-MFT-p-equation}
\end{align}
In the stationary case, one obtains
\begin{align}
\partial_x\left[ D(\n_u)\partial_x\n_u -\sigma(\n_u)(\partial_xp_u+u) \right]&=0;
\label{eq:SM-steady-MFT-1}
\\
D(\n_u)\partial_x^2p_u+\frac{\sigma'(\n_u)}{2}\left(\partial_xp_u+u\right)^2&=0,
\label{eq:SM-steady-MFT-2}
\end{align}
where $(\n_u,p_u)$ is a solution of the steady-MFT equations with the boundary conditions given above. The rescaled cumulant generating function is then
\begin{align}
{\cal F}(u)&:= \lim_{N\to\infty}N\Psi_N(u)= \lim_{N\to\infty}\lim_{\tau\to\infty}\frac{1}{N\tau }\log Z_{\tau}(u)
\nonumber\\
&=\int_0^1 dx\, {\cal H}_u[\n_u,p_u]= \int_0^1 dx\, \left[-D(\n_u)\partial_x\n_u(\partial_xp_u+u) +\frac{\sigma(\n_u)}{2}(\partial_xp_u+u)^2\right].
\label{eq:SM-current-CGF}
\end{align}
Its derivatives generate the rescaled asymptotic current cumulants,
\begin{align}
\frac{d^m{\cal F}}{du^m}\Big\vert_{u=0}&= \lim_{N\to\infty}\lim_{\tau\to\infty} \frac{1}{N\tau} \left\langle( {\cal Q}_{\tau})^m\right\rangle^c=\lim_{N\to\infty}\lim_{s\to\infty} \frac{N}{s}\left\langle (Q_s)^m\right\rangle^c.
\label{eq:SM-current-cumulants}
\end{align}
At $u=0$, the stationary solution is $p_0=0$ together with the density profile~\eqref{eq:SM-stationary-profile}, and Eq.~\eqref{eq:SM-current-CGF} gives ${\cal F}(0)=0$, as required. Note that for constant $D(\n)=D_0$, the ISEP reduces to the SSEP, for which a closed analytical expression for the cumulant generating function was obtained in Refs.~\cite{Derrida2004,derrida_2007}. The same expression gives the leading large-$N$ behavior of the QSSEP current cumulants, recently investigated in~\cite{Costa_emergence_universality_2026_PRL,albert2026universalclassicalquantumfluctuations}.\\

We finally discuss briefly the numerical procedure used to determine ${\cal F}(u)$. We first introduce the steady biased current
\begin{equation}\label{eq:SM-biased-curr}
\jj_u:=-D(\n_u)\partial_x\n_u+\sigma(\n_u)(\partial_xp_u+u),
\end{equation}
which is independent of $x$ because of Eq.~\eqref{eq:SM-steady-MFT-1}. The steady Hamiltonian density is also independent of $x$,
\begin{equation}
\partial_x{\cal H}_u=q_u\,\partial_x\jj_u-\partial_x\n_u\left[D(\n_u)\partial_xq_u+\frac{\sigma'(\n_u)}{2}q_u^2\right]=0,
\end{equation}
with shorthand $q_u:=\de_xp_u+u$. Indeed, the first term vanishes by Eq.~\eqref{eq:SM-steady-MFT-1}, while the term in square brackets vanishes by Eq.~\eqref{eq:SM-steady-MFT-2}. Since the system has unit size, Eq.~\eqref{eq:SM-current-CGF} implies that this constant is
\begin{equation}\label{eq:SM-K}
{\cal F}(u)={\cal H}_u[\n_u,p_u]=-D(\n_u)\partial_x\n_u(\partial_xp_u+u)+\frac{1}{2}\sigma(\n_u)(\partial_xp_u+u)^2.
\end{equation}
Eq.~\eqref{eq:SM-biased-curr} and \eqref{eq:SM-K} above then yield
\be
\left[D(\n_u)\partial_x\n_u\right]^2=\jj_u^2-2{\cal F}(u)\sigma(\n_u),
\ee
or equivalently, assuming a monotonic profile for $\n_u$,
\begin{equation}\label{eq:SM-MFT-biased-profile}
\partial_x\n_u = \frac{\epsilon}{D(\n_u)}\sqrt{\jj_u^2-2{\cal F}(u) \sigma(\n_u)},
\end{equation}
with branch specified by the reservoir imbalance, $\epsilon={\rm sgn}(n_b-n_a)$. Integrating Eq.~\eqref{eq:SM-MFT-biased-profile} between the two boundaries gives
\begin{equation}
1=\int_{n_a}^{n_b}dn\, \frac{D(n)} {\epsilon\sqrt{\jj_u^2-2{\cal F}(u)\sigma(n)}}.
\label{eq:SM-ISEP-dens-constraint}
\end{equation}
A second relation follows from the boundary conditions on $p_u$,
\begin{equation}
u=\int_0^1dx\,\left(\partial_xp_u+u\right)=\int_0^1 dx\, \frac{\jj_u+D(\n_u)\partial_x\n_u}{\sigma(\n_u)}.
\end{equation}
Substituting Eq.~\eqref{eq:SM-MFT-biased-profile} and changing the integration
variable from $x$ to $n=\n_u(x)$ then gives
\begin{equation}
u=\int_{n_a}^{n_b}\frac{dn}{\sigma(n)/D(n)} \left[1+\frac{\jj_u}{\epsilon\sqrt{\jj_u^2-2{\cal F}(u)\sigma(n)}}\right].
\label{eq:SM-ISEP-counting-constraint}
\end{equation}
Eq.~\eqref{eq:SM-ISEP-dens-constraint} and \eqref{eq:SM-ISEP-counting-constraint} directly determine $\jj_u$ and ${\cal F}(u)$ for any fixed value of $u$. They are therefore suitable for a numerical two-dimensional root-finding procedure initiated by the untilted solution
\begin{equation}
{\cal F}(0)=0, \quad \jj_0=\Phi(n_a)-\Phi(n_b),
\qquad
\Phi(n):=\int^n dq\,D(q).
\end{equation}
In the case of interest, the diffusivity and mobility are related by $\sigma(\n)=D(\n)\sigma_0(\n)$, with $\sigma_0(\n)=2\n(1-\n)$ the bare exclusion mobility, and the function $\Phi(n)$ is specified in Eq.~\eqref{eq:SM-Phi}.
\smsection{Numerical methods}
\label{app:numerics}
In this appendix, we describe how the tensor-network calculation is performed for the microscopic spin model \eqref{eq:DynamicsAve_rho}.  The
procedure can be briefly summarized in three distinct steps.  We first evolve the Lindblad
equation~\eqref{eq:DynamicsAve_rho} to its non-equilibrium steady state (NESS)
in a vectorized representation \cite{zwolak04,Jaschke_2019}.  We then measure the
magnetization and the current on the lattice and use  Fick's law~\eqref{eq:Fick} to find the local diffusivity.
Finally, we fit the estimated diffusivity to the form
in Eq.~\eqref{eq:D(n)}.  This fixes the parameters $D_0, \lambda$ and therefore the two effective couplings
$(\eta_{\rm eff},\Delta_{\rm eff})$ and hence, through
Eq.~\eqref{eq:mobility}, all the transport coefficients entering the MFT
predictions for the connected correlations, and the current cumulants.  We focus on the set of parameters $\varepsilon=-1 , \Delta=0.5 ,\eta=0.2, \gamma=1 , \mu=0.3$ for all the numerical results, in the main text as well as those presented in this section. At the end of this section, we present the convergence checks related to bond dimension, TEBD2 time step, and system size.

\smsubsection{Vectorized Liouvillian and quantum-number-preserving MPS}
\label{app:vectorisation}

As we already mentioned, the simulation is performed in the vectorised picture,
\begin{equation}
  \hat\rho=\sum_{a,b}\rho_{ab}|a\rangle\langle b|
  \quad\longmapsto\quad
  |\rho\rangle\!\rangle
  =\sum_{a,b}\rho_{ab}\,|a\rangle\otimes |b\rangle^{*},
  \label{eq:num_vectorisation}
\end{equation}
for which
$|\hat A\hat\rho\hat B\rangle\!\rangle
=(\hat A\otimes\hat B^{T})|\rho\rangle\!\rangle$. Consequently, the density matrix is represented as a vector in the doubled Hilbert space $\mathcal{H} \otimes \mathcal{H}^*$, with $\mathcal{H}$ being a $2^N$ dimensional Hilbert space.
$|\rho\rangle\!\rangle$ is stored as an MPS with physical legs of local dimension $q=4$, carrying a ket and a bra qubit configuration. 
Equation~\eqref{eq:DynamicsAve_rhotilted} consequently becomes
$\partial_t|\rho\rangle\!\rangle=\mathbb L\,|\rho\rangle\!\rangle$, with
\begin{equation}
 \label{eq:vecL}
\begin{aligned}
  \mathbb L={}&-i\bigl(\hat H_{\rm xxz}\otimes\hat{\mathbb I}
  -\hat{\mathbb I}\otimes (\hat{H}_{\rm xxz})^{T}\bigr)
   + \mathbb{L}_\eta + \mathbb{L}_{\rm bdy}
 \\  &\mathbb{D}_{\hat L_j}:=
  \hat L_j\otimes\hat L_j^{*}
  -\tfrac{1}{2}\hat L_j^{\dagger}\hat L_j\otimes\hat{\mathbb I}
  -\tfrac{1}{2}\hat{\mathbb I}\otimes
       \bigl(\hat L_j^{\dagger}\hat L_j\bigr)^{T} 
       \\ & \mathbb{L}_\eta := \eta \sum_{j=1}^N \mathbb{D}_{\hat \sigma^z_j} \quad,   \quad  \mathbb{L}_{\rm bdy}:=\!\!\sum_{\substack{p \in \left\{1,N\right\} \\
a\in\left\{ +,-\right\}
}}  \Gamma_{p,a} \mathbb{D}_{\hat \sigma^a_p}
\end{aligned}
\end{equation}
where $\mathbb{L}$, $\mathbb{D}_{\hat L_j}$,
$\mathbb{L}_{\eta}$, and $\mathbb{L}_{\rm bdy}$ denote the
superoperator forms of the Liouvillian and the dissipative contributions
of our microscopic model. In the doubled space, the dynamics possesses
a weak $U(1)$ symmetry~\cite{Buča_2012,Guo2025designingopen}, generated
by the ket--bra magnetization difference
\begin{equation}
 \hat Q_{\rm diff}
  =
  \sum_{j=1}^{N}
  \bigl(
  \hat\sigma^z_{j} \otimes\hat{\mathbb I}
  -
 \hat{\mathbb I} \otimes \hat\sigma^z_{j}
  \bigr).
  \label{eq:num_qn}
\end{equation}
because the jump terms $\hat L_{\hat \sigma^a_p}\otimes\hat L_{\hat \sigma^a_p}^{*}$ within the boundary dissipator $\mathbb{L}_{\rm bdy}$ change the bra and ket magnetization by the same amount, while all other terms trivially conserve bra and ket magnetization separately. Indeed, defining the simultaneous ket--bra rotation
\begin{equation}
  \mathbb U_\phi
  =
  e^{i\phi\hat Q_{\rm diff}},
  \qquad
  \mathbb U_\phi\mathbb L\mathbb U_\phi^\dagger
  =
  \mathbb L,
  \qquad\Longleftrightarrow\qquad
  [\mathbb L,\hat Q_{\rm diff}]=0,
\end{equation}
shows that the dynamics decomposes into sectors of fixed
$\hat Q_{\rm diff}$.
The symmetry is weak because only the combined 
ket--bra action is conserved, whereas the ket and bra magnetization
need not be conserved separately. The symmetry allows for the use of a quantum number-conserving MPS ~\cite{SciPostPhysCodeb.4,Bernier2018LightCone}, which renders the tensors into block-sparse form in the $\hat Q_{\rm diff}$
sectors.  Thus, contractions and
singular-value decompositions are performed within the allowed
blocks, which yields a substantial
reduction of both memory and runtime. The initial state of all the simulations is the infinite-temperature one;therefore subsequently means that we work in the sector of
$Q_{\rm diff}=0$.

For any operator $\hat O$,  expectation values and the trace are
obtained from the vectorized identity as
\begin{equation}
  \langle\hat O\rangle_t
  =\frac{\langle\!\langle\hat{\mathbb I}|
  (\hat O\otimes\hat{\mathbb I})|\rho(t)\rangle\!\rangle}
  {\langle\!\langle\hat{\mathbb I}|\rho(t)\rangle\!\rangle},
  \qquad
  \operatorname{tr}\hat\rho(t)
  =\langle\!\langle\hat{\mathbb I}|\rho(t)\rangle\!\rangle.
  \label{eq:num_expectation}
\end{equation}

The time evolution is performed according to a TEBD2 scheme, which can be briefly described as follows: we split the nearest-neighbor terms into
commuting even- and odd-bond layers,
$\mathbb L=\mathbb L_{\rm even}+\mathbb L_{\rm odd}$, whose terms commute within each layer. We assign each boundary
dissipator to the layer containing the adjacent boundary bond.  One time step is
implemented with the symmetric second-order Suzuki--Trotter
formula
\begin{equation}
  e^{\delta t\,\mathbb L}
  =e^{\frac{\delta t}{2}\mathbb L_{\rm even}}
   e^{\delta t\,\mathbb L_{\rm odd}}
   e^{\frac{\delta t}{2}\mathbb L_{\rm even}}
   +\mathcal O(\delta t^3).
  \label{eq:trotter}
\end{equation}
The error per step is $\mathcal O(\delta t^3)$, so that the accumulated error at
fixed time is $\mathcal O(\delta t^2)$.  Since the time evolution generator $\mathbb{L}$ is time
independent, the gates are built once and reused at every step.  After applying a two-site gate, the MPS is compressed by discarding singular values below the
cutoff $10^{-13}$ and by imposing the bond dimension $\chi$, which in our numerics takes the values $120$ or $200$.

Numerical calculations for the tilted model Eq.~\eqref{eq:DynamicsAve_rhotilted} are performed in the same way.

\smsubsection{Effective parameters and MFT results}
\label{app:observables}

The simulation is performed in the Pauli basis.  Densities and their cumulants
are recovered from $\hat n_j=(\hat{\mathbb I}+\hat\sigma^z_j)/2$, so that
$
  \bar n_j=\frac{1+\langle\hat\sigma_j^z\rangle}{2}
  \quad\Longleftrightarrow\quad
  \langle\hat\sigma^z_{j=xN}\rangle=2\bar n_x-1 ,
  \label{eq:num_density}
$
and connected correlations from
\begin{equation}
  \langle\hat n_i\hat n_j\rangle^{c}
  =\tfrac{1}{4}\langle\hat\sigma_i^z\hat\sigma_j^z\rangle^{c},
  \qquad
  \langle\hat n_i\hat n_j\hat n_k\rangle^{c}
  =\tfrac{1}{8}\langle\hat\sigma_i^z
  \hat\sigma_j^z\hat\sigma_k^z\rangle^{c}.
  \label{eq:num_cumulant_conversion}
\end{equation}

The particle current in the bulk of our microscopic model follows from the discrete continuity equation $\partial_t\langle\hat n_i\rangle
=\langle\hat\jmath_{i-1}\rangle-\langle\hat\jmath_i\rangle$ which is actually the Heisenberg equation $\partial_t \expval{\hat n_i}= i \expval{[\hat H_{\rm xxz},\hat n_i]}$. The
anisotropy and the dephasing terms of Eq.~\eqref{eq:DynamicsAve_rho} commute
with $\hat n_i$ and therefore do not contribute in the bulk, so that one recovers
\begin{equation}
  \hat\jmath_i=-i[\hat h_{i,i+1}, \hat n_i]
  =\varepsilon\left(
  \hat\sigma_i^x\hat\sigma_{i+1}^y
  -\hat\sigma_i^y\hat\sigma_{i+1}^x\right)
  =2i\varepsilon\left(
  \hat\sigma_i^+\hat\sigma_{i+1}^-
  -\hat\sigma_i^-\hat\sigma_{i+1}^+\right).
  \label{eq:num_current}
\end{equation}
for $j=1, \dots, N-1$.
In the continuum limit, the diffusive current (as in  Eq.~\eqref{eq:effec-Current-contlimit}) satisfies  $\bar J_x= \lim_{N \to \infty}N\langle \hat j_{j=xN} \rangle= - D(\bar n_x) \nabla \bar n_x$ with $\nabla = \partial_x$. This relation is used to recover the local diffusivity $D(\bar n_x)$. Specifically, we scan through our system with a window of width $w$, which we denote as $W_j=\{\frac{j}{N},\frac{j+1}{N},\dots \frac{j+w-1}{N} \}$. Over that window, we find the average density $\bar n_{W_j}=w^{-1} \sum_{x \in W_j} \bar n_{x} $ and  current $\langle j_{W_j} \rangle =w^{-1} \sum_{x \in W_j} \langle j_{j=xN}\rangle $  and finally the gradient of $ \nabla \bar n_{W_j}$ via linear fitting. Then, we extract $D(\bar n_{W_j})=-N\expval{j_{W_j}}/\nabla \bar n_{W_j}$.

\begin{figure*}
\centering

\begin{overpic}[width=0.34\textwidth]
    {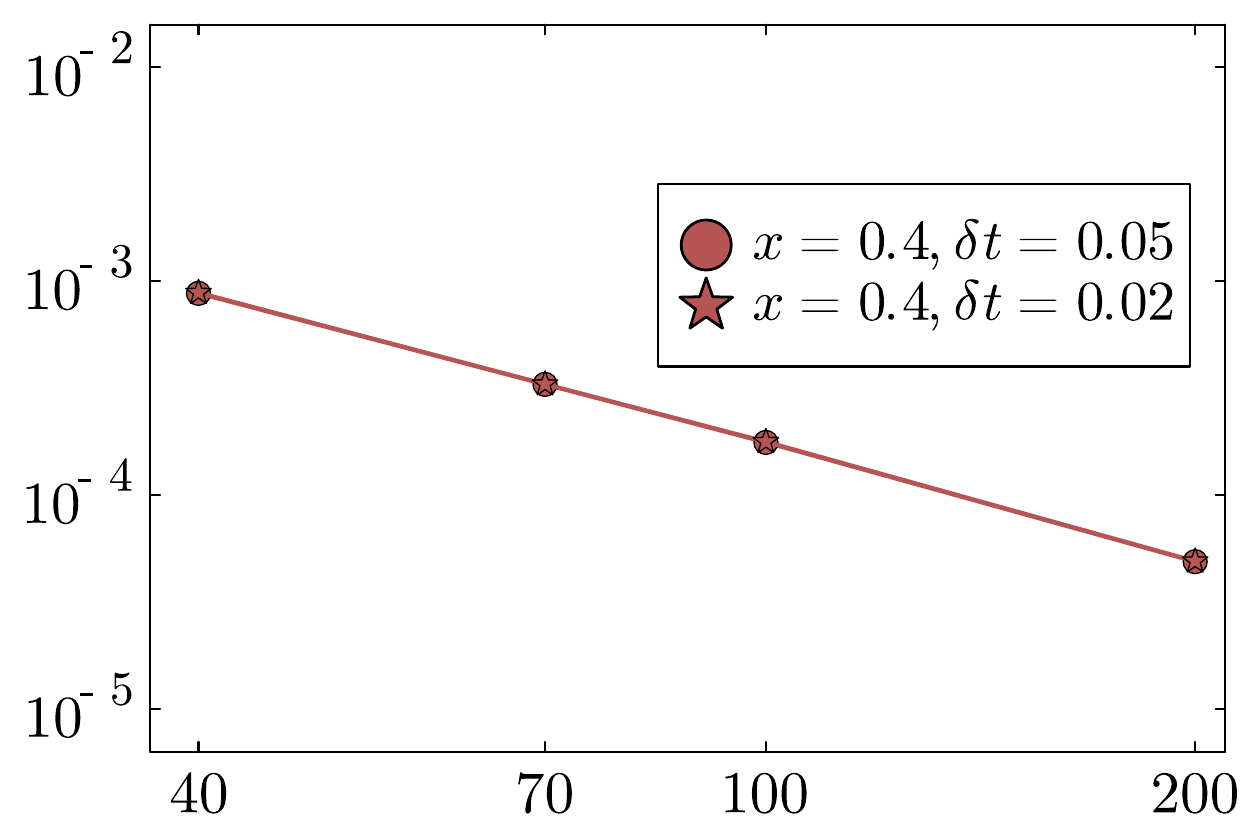}
    \put(50.3,58){(g)}
    \put(50,-4){$N$}
    \put(-7,34){\rotatebox[origin=c]{90}
        {$\left|
        \langle\hat\sigma^z_{N/2}
        \hat\sigma^z_{N/6}
        \hat\sigma^z_{i=xN}\rangle^{c}
        \right|$}}
\end{overpic}
\hspace{8mm}
\begin{overpic}[width=0.34\textwidth]
    {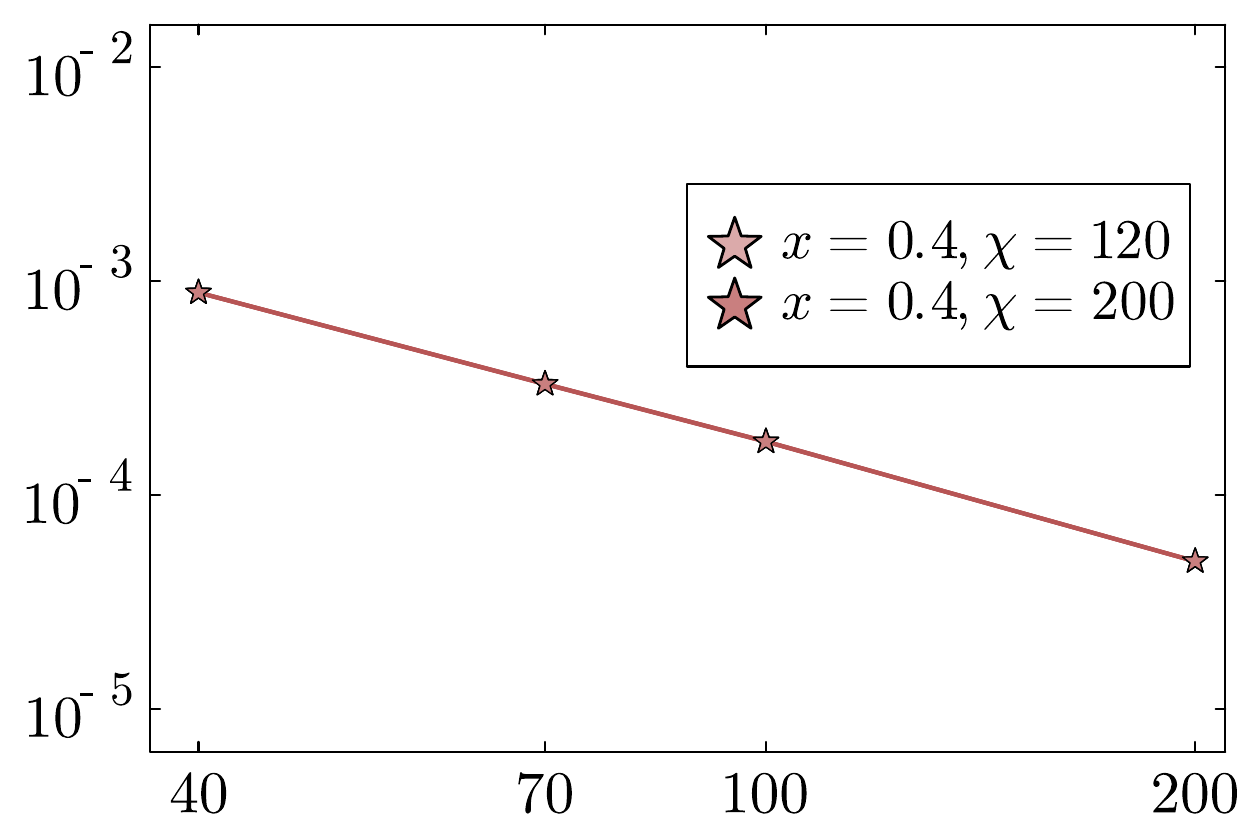}
    \put(50.3,58){(h)}
    \put(50,-4){$N$}
    \put(-7,34){\rotatebox[origin=c]{90}
        {$\left|
        \langle\hat\sigma^z_{N/2}
        \hat\sigma^z_{N/6}
        \hat\sigma^z_{i=xN}\rangle^{c}
        \right|$}}
\end{overpic}
\caption{
Convergence checks for the tensor-network results used in the main
text. The microscopic parameters are
$\varepsilon=-1$, $\eta=0.2$, $\Delta=0.5$, $\gamma=1$ and $\mu=0.3$,
with SVD cutoff $10^{-13}$.
The left column tests the TEBD2 time-step dependence using
$\delta t=0.05$ and $0.02$ at fixed $\chi=200$, whereas the right column
tests the MPS-truncation dependence using
$\chi=120,200$ at fixed $\delta t=0.02$. Moreover, the markers  $\opencircle, \openstar$ are used to indicate numerical data for $\delta t= 0.05 , 0.02$ respectively.
(a),(b) Effective couplings
$(\Delta_{\rm eff},\eta_{\rm eff})$, extracted from the fit of the local
diffusivity in Eqs.~\eqref{eq:D_fit}--\eqref{eq:eff_params} and their flow with system size to their RG fixed point.
(c),(d) Rescaled connected two-point spin correlation
$4C_{x,1/2}$, with one position fixed at $y=1/2$ for $N=200$; only the regular
non-equilibrium part is shown.
(e),(f) Rescaled current cumulant generating function
$\mathcal{F}(u)$ at $N=200$.
(g),(h) Absolute value of the connected three-point spin
correlation for the rescaled positions indicated in the legends.
The good agreement between results obtained with different time steps and bond dimensions demonstrates the convergence of our numerical results with respect to both the Trotter step and the bond dimension.
}
\label{fig:convergence}
\end{figure*}

The data are used to perform the following fit 
\begin{equation}
  D(\bar n)=A+B\,\bar n(1-\bar n),
  \label{eq:D_fit}
\end{equation}
related to Eq.~\eqref{eq:D(n)} via
\begin{equation}
  A=D_0=\frac{2\varepsilon^2}{\eta_{\rm eff}} \quad ,
  \quad
  B=2D_0\lambda,
  \label{eq:num_AB_match}
\end{equation}
Inverting Eq.~\eqref{eq:num_AB_match} one recovers  $D_0,\lambda $ and then the relation
$\lambda=-\bigl[1+(\eta_{\rm eff}/\Delta_{\rm eff})^{2}\bigr]^{-1}$ gives the
effective couplings,
\begin{align}
  \eta_{\rm eff}&=\frac{2\varepsilon^{2}}{A},
  \nonumber\\
  \Delta_{\rm eff}&=
  \frac{2\varepsilon^{2}}{A}
  \sqrt{-\frac{B}{2A+B}} ,
  \label{eq:eff_params}
\end{align}
The intercept corresponds to the non-interacting limit of diffusivity $A=D_0$ and thus that of the SSEP, while $B$
encodes the leading interaction correction. In our case, we choose $w=12$, but in general any choice of $2<w \ll N$ causes very small differences in our results, which vanish in the thermodynamic limit.  Repeating the fit at each $N$ allows us to find the flow of $(\eta_{\rm eff},\Delta_{\rm eff})$ towards the thermodynamic limit, where the effective description is given by ISEP.
As we demonstrate in Fig. ~\ref{fig:convergence}(a,b), by increasing the system size to $N\le 200$, we obtain $\eta_{\rm eff}, \Delta_{\rm eff}$ that are well converged to the thermodynamic values and can capture with good accuracy the fixed point of the RG flow for the transport coefficients $D(n),\sigma(n)$. 

For the purpose of recovering the MFT predictions of correlations and CGF, we thus use the values of $\eta_{\rm eff},\Delta_{\rm eff}$  at $N=200$.  The mobility of ISEP is then fixed by the fluctuation-dissipation relation Eq.~\eqref{eq:mobility} and the magnetization profile via Eq.~\eqref{eq:theoretical-density-profile}. To obtain the two-point correlations $C_{x,y}$, we numerically solve Eq.~\eqref{eq:eq-for-C} and find the stationary solution,
 with the Dirichlet conditions
$C_{0,y}=C_{1,y}=C_{x,0}=C_{x,1}=0$ and the analytical solution of the transport coefficients
$D(\bar n),\sigma (\bar n)$. Fig.~\ref{fig:convergence}(c,d) shows a numerical
comparison between tensor-network results on $N=200$ and the ISEP prediction on distinct sites; the theoretical curve shown in
Fig.~\ref{fig:ISEPnumerical}(c,d) is the regular non-equilibrium part $C_{x,y}^{\rm neq}$, and the contact term
at $x=y$ is not included. The same transport coefficients $\eta_{\rm eff},\Delta_{\rm eff}$ are used to find the rescaled CGF $\mathcal{F}(u)$ based on the procedure described in  Sec.~\ref{sec:SM-current-MFT}. Fig.~\ref{fig:convergence}(e,f) supports reasonable agreement between the tensor-network and ISEP result for this quantity as well. 
\smsubsection{Convergence check}
\label{app:convergence}
In this final part, we present additional numerical results to corroborate the convergence in $\chi, \delta t$ and $N$. Specifically, we 
perform our simulations for two different time steps $\delta t=0.05,0.02$ and $\chi=120,200$ to verify that the numerical errors due to Trotterization and MPS truncation, respectively, are small enough and the data are well converged. We show this in Fig.~\ref{fig:convergence} for the effective parameters, rescaled CGF, and two- and three-point connected correlations.
\\
\begin{figure*}[h]
\begin{overpic}[width=0.34\textwidth]
    {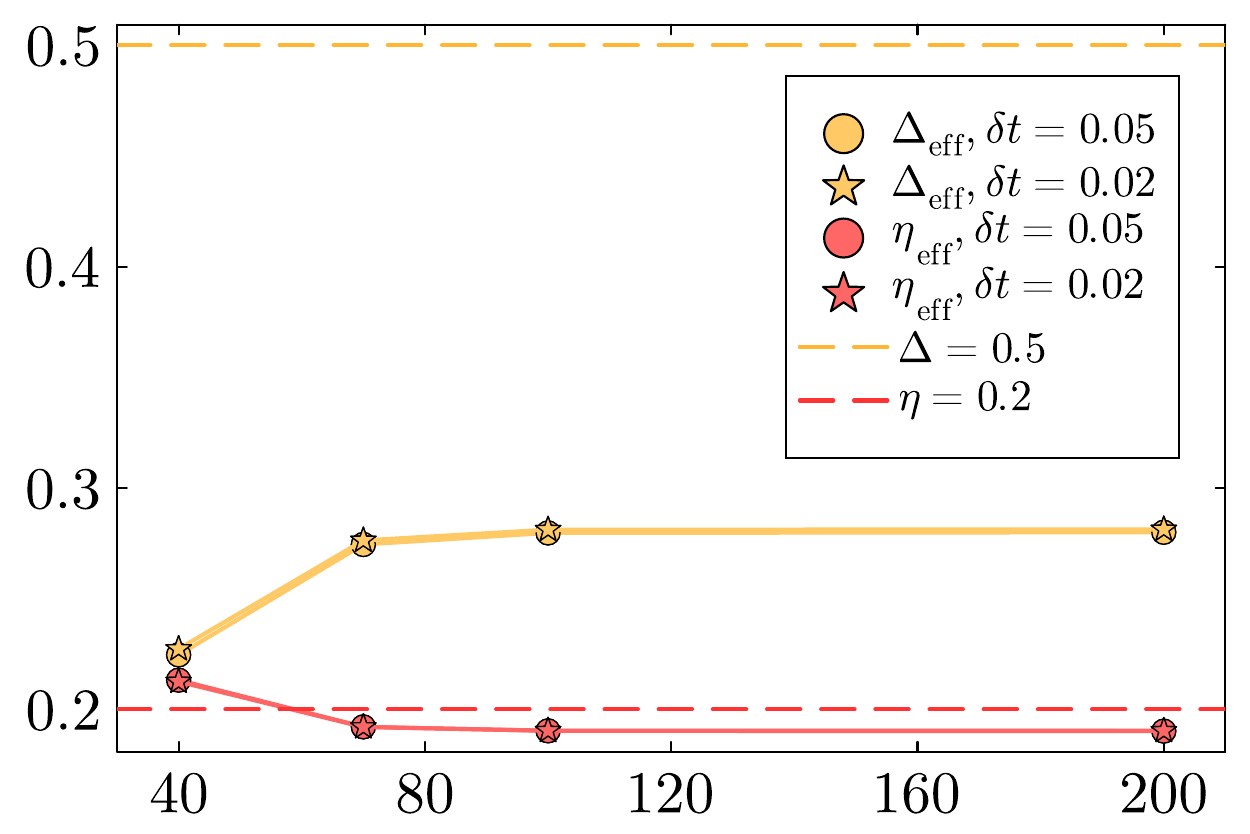}
    \put(50.3,58){(a)}
    \put(50,-4){$N$}
    \put(-6,34){\rotatebox[origin=c]{90}
        {$\Delta_{\rm eff},\,\eta_{\rm eff}$}}
\end{overpic}
\hspace{8mm}
\begin{overpic}[width=0.34\textwidth]
    {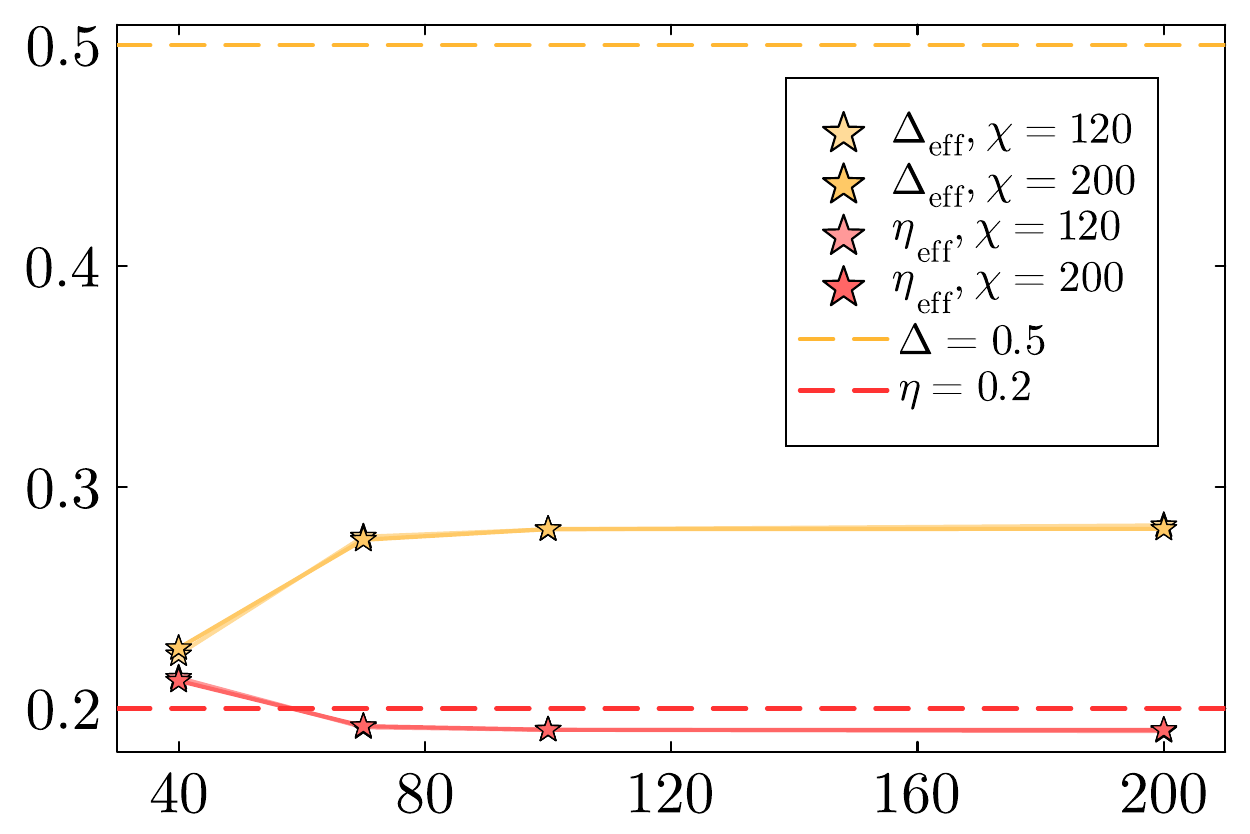}
    \put(50.3,58){(b)}
    \put(50,-4){$N$}
    \put(-6,34){\rotatebox[origin=c]{90}
        {$\Delta_{\rm eff},\,\eta_{\rm eff}$}}
\end{overpic}

\vspace{0.85cm}


\begin{overpic}[width=0.34\textwidth]
    {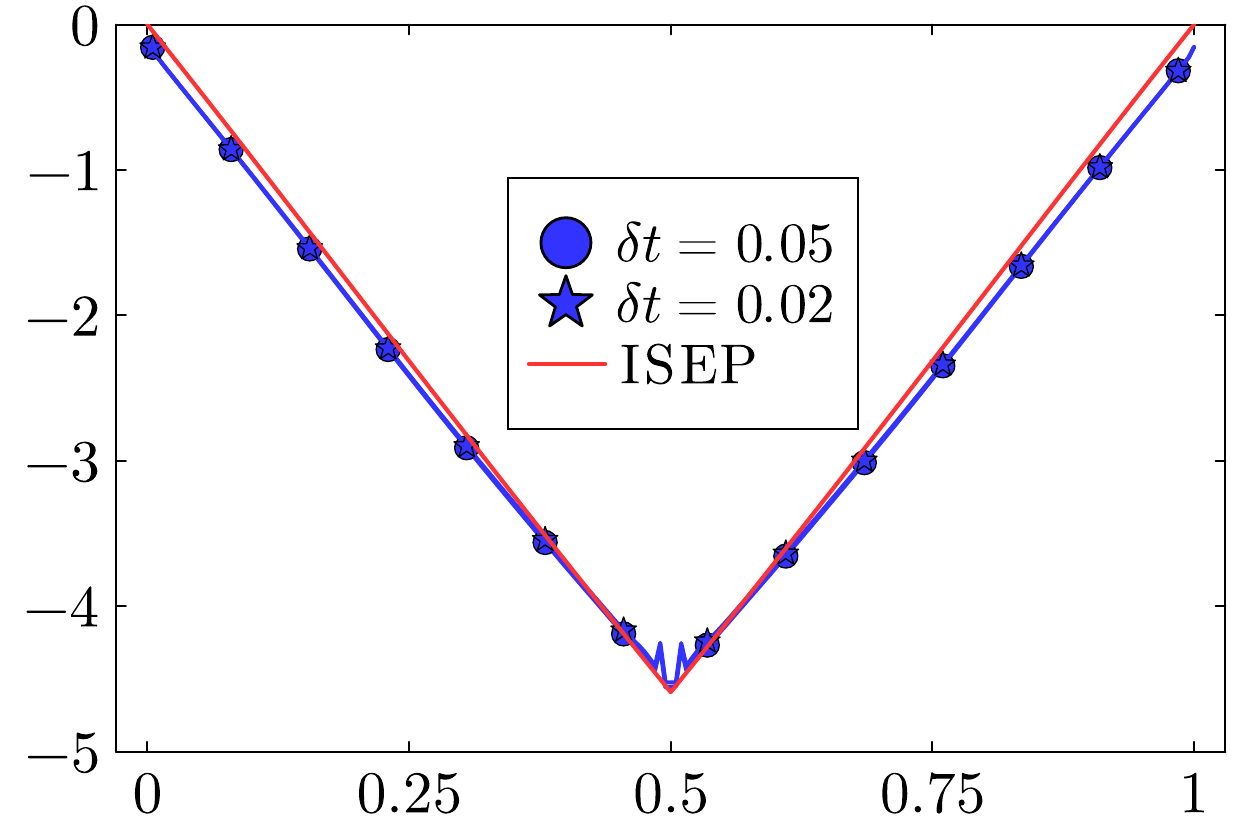}
    \put(50.3,58){(c)}
    \put(51,-4){$x$}
    \put(-7,34){\rotatebox[origin=c]{90}
        {$4C_{x,1/2}\times 10^{2}$}}
\end{overpic}
\hspace{8mm}
\begin{overpic}[width=0.34\textwidth]
    {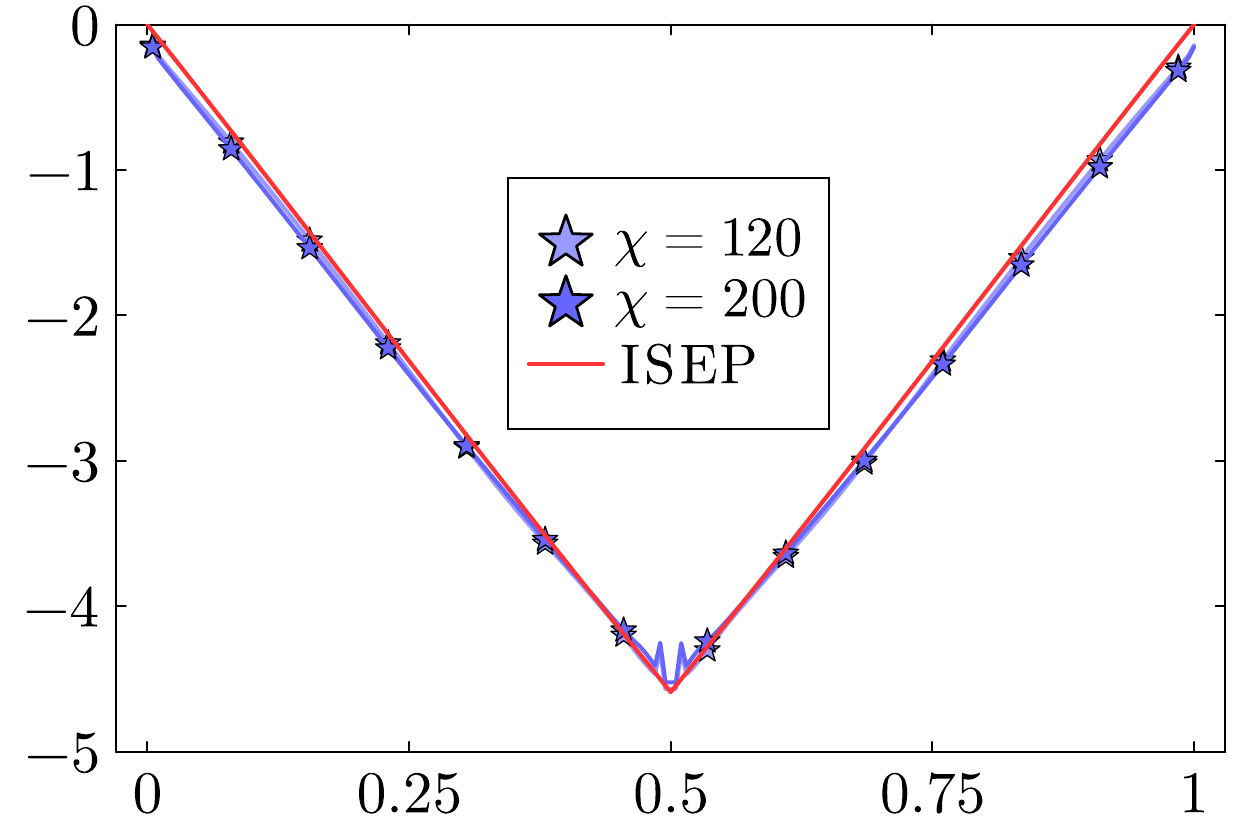}
    \put(50.3,58){(d)}
    \put(51,-4){$x$}
    \put(-7,34){\rotatebox[origin=c]{90}
        {$4C_{x,1/2}\times 10^{2}$}}
\end{overpic}

\vspace{0.85cm}


\begin{overpic}[width=0.34\textwidth]
    {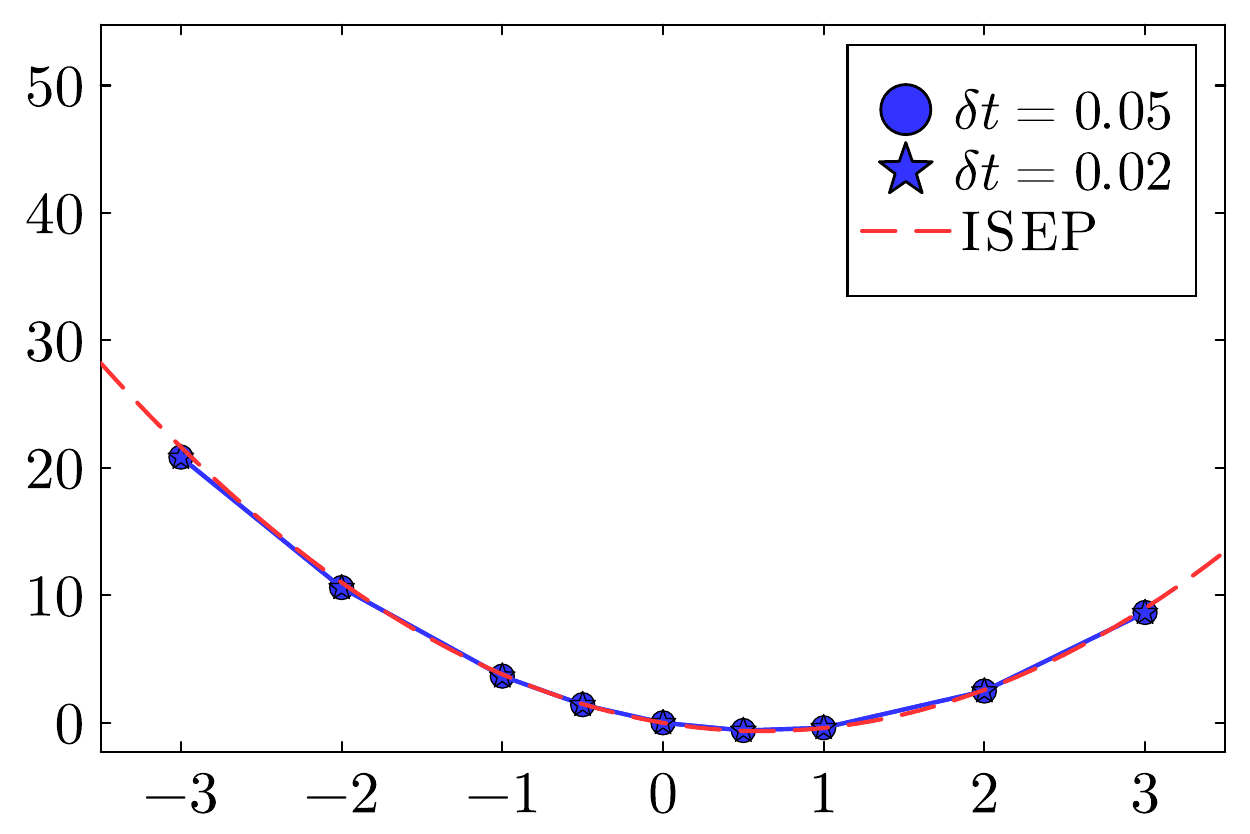}
    \put(50.3,58){(e)}
    \put(51,-4){$u$}
    \put(-6,34){\rotatebox[origin=c]{90}
        {$\mathcal{F}(u)$}}
\end{overpic}
\hspace{8mm}
\begin{overpic}[width=0.34\textwidth]
    {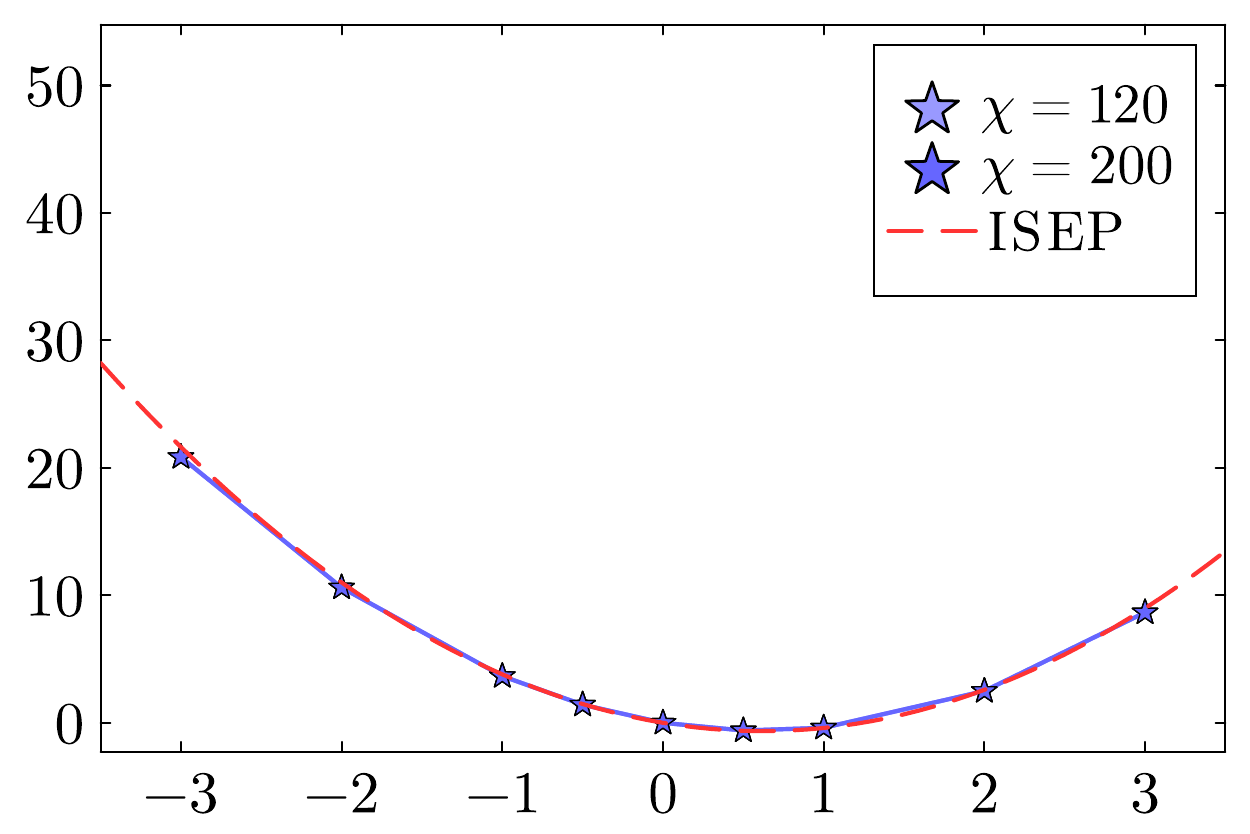}
    \put(50.3,58){(f)}
    \put(51,-4){$u$}
    \put(-6,34){\rotatebox[origin=c]{90}
        {$\mathcal{F}(u)$}}
\end{overpic}

\end{figure*}

\end{document}